\documentclass[conference]{IEEEtran}
\pdfoutput=1
\usepackage{graphicx}
\usepackage{amsthm}
\usepackage{amssymb}
\usepackage{amsmath}
\usepackage{balance}
\usepackage{enumerate}
\begin{document}


\title{Decentralizing MVCC by Leveraging Visibility}


\author{
\IEEEauthorblockN{Xuan Zhou}
\vspace{3mm}
\IEEEauthorblockA{School of Data Science and Engineering\\
East China Normal University\\
3663 Zhongshan North Road\\
Shanghai 200062, China\\
Email: xzhou@dase.ecnu.edu.cn}
\and
\IEEEauthorblockN{Xin Zhou \hspace{2mm} Zhengtai Yu \hspace{2mm} Hua Guo }
\vspace{3mm}
\IEEEauthorblockA{DEKE Lab\\
Renmin University of China\\
59 Zhongguancun Street\\
Beijing 100872, China\\
Email: guohua2016@ruc.edu.cn}
\and
\IEEEauthorblockN{Kian-Lee Tan}
\vspace{3mm}
\IEEEauthorblockA{School of Computing\\
National University of Singapore\\
13 Computing Dr.\\
Singapore 117417\\
Email: tankl@comp.nus.edu.sg}}

\IEEEtitleabstractindextext{%
\begin{abstract}
Multiversion Concurrency Control (MVCC) is a widely adopted concurrency control mechanism
in database systems, which usually utilizes timestamps to resolve
conflicts between transactions.
However, centralized allocation of timestamps is a potential bottleneck
for parallel transaction management. This bottleneck is becoming
increasingly visible with the rapidly growing degree of parallelism of
today's computing platforms. This paper introduces Visibility based Concurrency Control (ViCC),
a series of CC mechanisms that allow transactions to
determine their timestamps autonomously, without relying on
centralized coordination. As such, ViCC can scale well,
rendering it suitable for various multicore and MPP platforms.
Extensive experiments are conducted to demonstrate its advantage over existing approaches.
\end{abstract}

}

\maketitle


%

\IEEEraisesectionheading{\section{Introduction}\label{sec:introduction}}

\IEEEPARstart{T}{he} application of Multiversion Concurrency Control (MVCC) in real world systems is pervasive. A significant number of the most popular database systems, including Oracle \cite{oracleSI}, SQL Server \cite{SQLServerSI}, PostgreSQL \cite{ports2012serializable}, adopt some kinds of MVCC as their concurrency control schemes. The main advantage of MVCC is at its capability to avoid blocking caused by read-write conflicts \cite{berenson1995critique}. This gives it tremendous performance gain in read intensive applications. In recent years, various mechanisms of MVCC have been proposed and adopted by Web-scale applications~\cite{corbett2013spanner}. This further confirmed its suitability in transaction management of modern applications.

MVCC usually relies heavily on timestamps to determine the temporal relationship among transactions, which allows it to detect conflicts causing problem of data consistency. However, such temporal relationship is, by definition, based on a single clock. To allocate timestamps from a single clock, a central coordinator seems indispensable. As a result, the coordinator can become a potential bottleneck to scalability or a single point of failure for the entire system.

Many of today's computing platforms come with a high degree of parallelism.
On the one hand, due to the weakening of Moore's Law \cite{schulz1999end,kish2002end}, CPU manufacturers have started adding more and more cores to a single chip to enhance its processing power. If on-chip parallelism keeps increasing, a server with hundreds of cores will be common in the foreseeable future \cite{pavlo15}. To prepare for the architectural shift, a number of research projects have recently been launched, aiming to architect database systems on hundreds to thousands of cores \cite{yu2014}. On the other hand, as the data volume in our IT infrastructures grows exponentially, scalability in a large computer cluster is regarded as an important capability of modern database systems. A new generation of parallel database systems, such as those classified as NoSQL \cite{cattell2011scalable} or NewSQL \cite{stonebraker2012newsql} databases, have been invented to support such scalability. On a highly parallelized platform, centralized coordination is not desirable, for it may severely impair the scalability and fault tolerance of the system. This prompts
us to rethink the design of MVCC mechanisms.

\begin{figure}
\centering
\includegraphics[width=.4\textwidth]{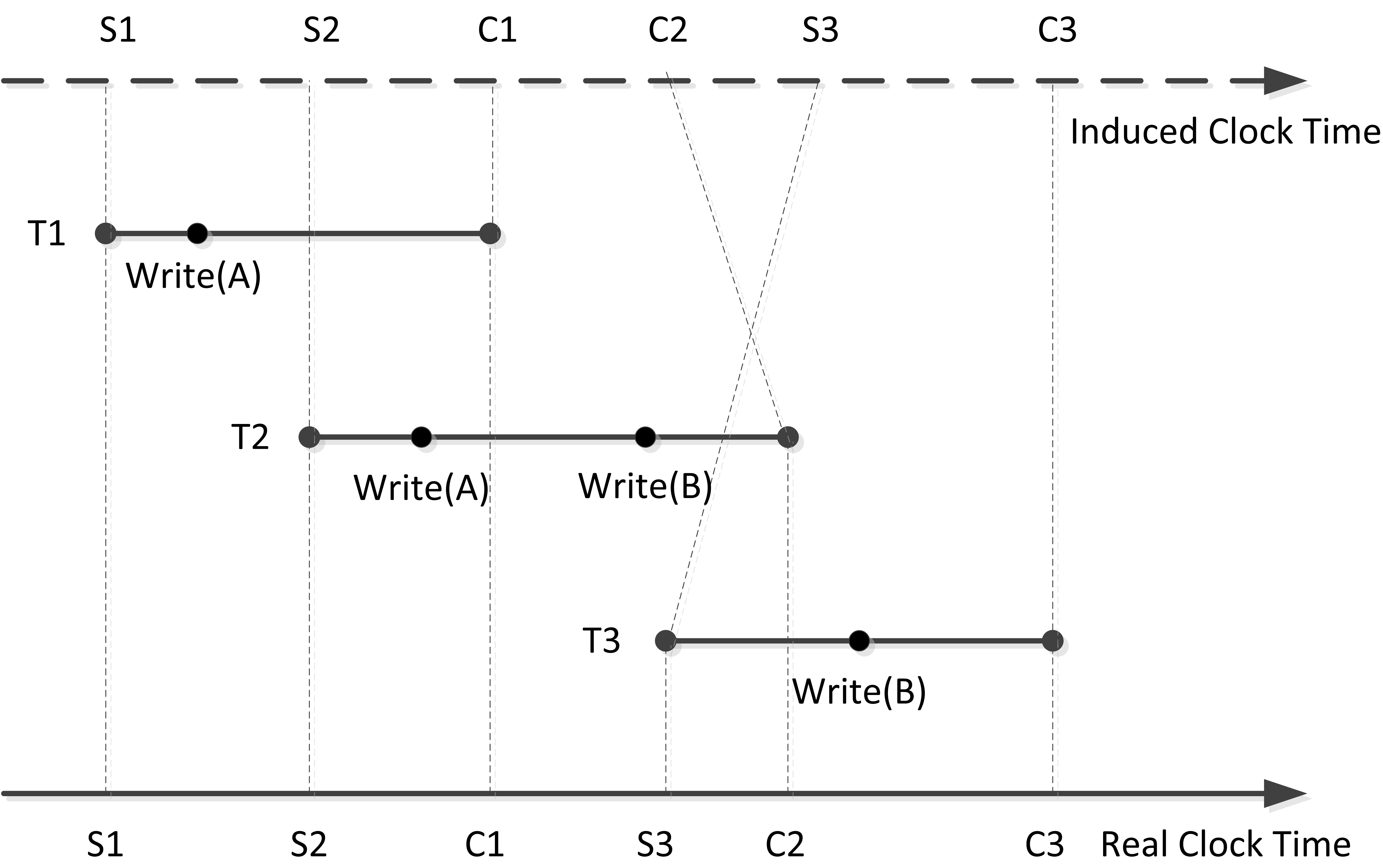}
\caption{Determining Timestamps Post-priori (A is the only data shared by T1 and T2. B is the only data shared by T2 and T3.)}
\label{fig:timeline}
\vspace{0mm}
\end{figure}

In this paper, we show that MVCC can be implemented efficiently without any centralized coordination. The key idea is to let transactions to negotiate about and determine their timestamps autonomously.

Snapshot Isolation (SI) is a popular MVCC mechanism. Under SI, each transaction is allocated with two timestamps, one at its beginning (known as start time) and the other at its end (known as commit time).  If the time intervals of two transactions overlap, they are not allowed to modify the same piece of data. In Figure~\ref{fig:timeline}, $t_1$ and $t_2$ conflict, as they both attempt to update Item $A$ simultaneously. This simultaneity can be judged from their timestamps --
as $s_1<s_2<c_1$, the time intervals of $t_1$ and $t_2$ are deemed to overlap.
To ensure the isolation, we have to abort either $t_1$ or $t_2$.
However, this constraint imposed by traditional SI can be overly restrictive.
In fact, it is sometimes safe to allow overlapping transactions to modify the same data. Consider the case of $t_2$ and $t_3$, both attempting to modify $B$.
While the intervals of $t_2$ and $t_3$ overlap too, when $t_3$ starts to access $B$, $t_2$ has already committed. If the two transactions do not share any other data, $t_3$ can safely overwrite the version of $B$ generated by $t_2$, without harming the consistency of data. In other words, $t_3$ can be regarded as a transaction that starts after the commit of $t_2$. To achieve this, we need to manipulate the timestamps, i.e., $c_2$ and $s_3$, to force $c_2<s_3$.

Instead of allocating timestamps promptly from a real clock, is it possible to determine the timestamps in the aftermath of transaction execution and induce a
logical clock from the timestamps? (An induced clock is shown at the top of Figure~\ref{fig:timeline}.) The idea of Visibility based Concurrency Control (ViCC) was inspired by this thought. ViCC utilizes the concept of \emph{visibility}, a binary order among transactions. In ViCC, transactions determine their timestamps during or after execution, by consulting their visibility orders. This allows ViCC to get rid of the central clock for coordination, making it superior to traditional MVCC in scalability and fault tolerance (without a central point of failure). In this paper, we introduce three isolation levels of ViCC -- Consistent Visibility (CV), Posterior Snapshot Isolation (PostSI) and Serializable Visibility (SV). The latter two are in effect identical to traditional SI and Serializability. We discuss how to implement these isolation levels efficiently over MPP databases and conduct extensive evaluation to characterize their performance and demonstrate their advantage over alternative approaches.

In a preliminary version of this paper \cite{zhou2017}, we introduced CV and PostSI. In this extended version, we show that serializability can also be enforced efficiently by utilizing visibility. Additional experiments were conducted to evaluate it.

The rest of the paper is organized as follows. Section 2
reviews the related work. Section 3 introduces the concept of visibility and the three isolation levels based on it.
Section 4 presents the schedulers to enforce the isolation levels.
Section 5 discusses how to implement the schedulers on a shared-nothing architecture.
Section 6 reports our experimental results. Section 7 concludes
the paper and discusses directions for future work.

\section{Related Work}

Decentralized concurrency control has been studied extensively in the context of distributed or parallel databases \cite{Bernstein:1981:CCD,weikum2001transactional,Harding:2017:EDC}. Most systems resort to locking for distributed concurrency control \cite{Bernstein:1981:CCD,DBLP:conf/usenix/CowlingL12}, as locking seems relatively easy to decentralize. To decentralize a locking based concurrency controller, such as 2PL, we maintain a lock table on each data node, such that locking can be performed locally. Dead lock detection is usually necessary for lock based approaches. While distributed dead lock detection requires no centralized coordination, it can be expensive. Similarly, mechanisms of Optimistic Concurrency Control (OCC) are not difficult to decentralize either \cite{Thomasian:1998:DOC,Mu:2014:EMC} -- the read and write sets of a transaction can be partitioned and stored locally, and the validation can be conducted separately on each node. However, traditional 2PL and OCC based approaches do not share the advantage of MVCC. They either block or abort transactions when encountering read-write conflicts.

When considering MVCC, decentralization of concurrency control becomes a challenge, as most existing implementations of MVCC rely on timestamps to determine the right data version for a transaction to access.  
(While MV2PL \cite{Bernstein:1986:CCR:17299} does not require timestamp, it can only delay rather than avoiding blocking when confronted with read-write conflicts.)
In the literature, several approaches of distributed or parallel MVCC have been proposed \cite{schenkel2000federated,binnig2014distributed,sovran2011transactional}. They either aim to improve the scalability of distributed MVCC \cite{zhang2010supporting,zhang2011hbasesi,lee2013sap,binnig2014distributed,du2013clock}, or to provide high-availability support  \cite{sovran2011transactional,ardekani2013non,ardekani2013non,tripathi2015scalable}. However, most of the approaches still use central clocks to allocate timestamps. In what follows, we briefly review the work that attempts to alleviate centralized timestamp allocation.

In \cite{binnig2014distributed}, the authors introduced Distributed Snapshot Isolation (DSI), an SI scheme for MPP databases. They proposed four methods to implement DSI. Among the four, the \emph{incremental snapshot method} is regarded as the most efficient. 
In this method, a local transaction only interacts with the local clock to retrieve timestamps.
Only when a transaction attempts to access the data on a remote node, does it interact with that node to obtain an appropriate remote timestamp. To ensure the validity of remote timestamps, a global clock is still required, and a mapping between each local clock and the global clock is maintained. Each node communicates with the coordinator occasionally to keep the mapping up-to-date. Although this method can avoid centralized coordination for single-node transactions, it is still mandatory for cross-node transactions. Compared to DSI, ViCC eliminates the need for centralized coordination completely. 

To avoid using a central clock, another viable approach is to use synchronized distributed physical clocks (a.k.a. true time devices). A typical example is Spanner \cite{corbett2013spanner}, a distributed database system developed by Google. Spanner utilizes GPS clocks and atomic clocks to constraint the deviation among different physical clocks within a small error bound. It then builds its concurrency control mechanism upon the timestamps generated by the true time devices. However, as GPS clocks and atomic clocks are not common hardware, the approach of Spanner does not seem to be widely applicable. Instead of using hardware of high accuracy, Clock-SI \cite{du2013clock} resorts to an algorithmic approach that derives timestamps from loosely synchronized physical clocks. Loose clock synchronization \cite{adya1995efficient} would unavoidably result in skew of time. To deal with time skew, Clock-SI has to let a node falling behind to see only old data snapshots or to force an ahead node to wait for a behind node. This makes Clock-SI unstable, as enlarged clock skew will result in severe performance loss. 
ViCC chooses not to deal with synchronized physical clocks.

ViCC's approach is to determine timestamps at the end of each transaction, after the transaction's relationship with the other transactions is set in stone. This allows a transaction to set its own timestamp autonomously. The same idea has been shared by other research work. To the best of our knowledge, Timestamp-range Conflict Manager (TCM) proposed by Lomet et al. \cite{lomet2012multi} was one of the earliest attempts at adjusting timestamps during the conduction of transactions. It assigns each transaction with a single timestamp to determine its serial order. By adjusting the timestamps on the fly, it allows some conflicting transactions to proceed concurrently over different versions of the data.
As TCM's main purpose to improve the degree of concurrency, it did not consider decentralizing its mechanism over a distributed database.
TicToc \cite{yu2016tictoc} applies this idea to avoid centralized coordination on multicore platforms. TicToc combines OCC and MVCC to enforce serializability. It marks each data item with two timestamps to represent its valid period. The serial order of a transaction is then decided by synchronizing the time stamps of its read and write sets. The isolation levels enabled by ViCC is not limited to serializability. It enables two extra levels of isolation -- VC and PostSI.
While the methods of TCM and TicToc can potentially be extended to work with SI or Read Committed, these features remain insufficiently explored. Moreover, as they are mainly designed for single machines, it is unclear if their designs fit well with a distributed or parallel database. Detailed comparison between ViCC and TicToc can be found in Sections~5.2 and 6.3.

With the prevalence of multicore processors, some recent work \cite{Faleiro:2015:RSM,Neumann:2015:FSM,Dashti:2017:TRM} has studied how to scale MVCC on multicore platforms. In \cite{Faleiro:2015:RSM}, a unique MVCC mechanism named BOHM is proposed. It determines the versions of transactions¡¯ writes prior to their execution, so as to improve the parallelism of transaction processing. In \cite{Neumann:2015:FSM}, the authors proposed a carefully engineered MVCC mechanism which use Precise Locking to achieve enhanced performance. In \cite{Dashti:2017:TRM}, a transaction repairing scheme is introduced to speedup the ``abort and restart" phase of transactions. Nevertheless, none of the these approaches aims to get rid off centralized timestamp allocation.
While ViCC mainly consider distributed and parallel databases, it can potentially be applied to multicore platforms too.

Replication is commonly applied to to enhance the availability of a database. In \cite{elnikety2005database}, Elnikety et al. propose Generalized SI, which allows a transaction to push its start time earlier to facilitate concurrency control on replicated data.
In \cite{sovran2011transactional}, Sovran et al. propose Parallel Snapshot Isolation (PSI), a weaker isolation level than SI that allows different nodes to have different commit orderings. Using asynchronous commit orderings, PSI was shown to achieve significant performance improvement. In \cite{ardekani2013non}, an even weaker version of SI called non-monotonic SI was proposed for replicated databases. As non-monotonic SI further relaxes some constraints of PSI, it outperforms PSI in certain circumstances. Other related work on implementing SI over replicated databases can be found in \cite{daudjee2006lazy,jung2011serializable,lin2009snapshot,chairunnanda2014confluxdb,tripathi2015scalable}. In this paper, we do not consider data replication. The issue of data replication is actually orthogonal to that of timestamp allocation. Rather than being our competitors, these approaches are complementary to our work.

\section{Visibility based Isolation}
\subsection{The Concept of Visibility}

Visibility is a type of binary relationship among transactions.

\newtheorem{definition}{Definition}
\begin{definition}
\emph{(Visibility)}:
Let $t_i$ and $t_j$ be two transactions. We say that $t_i$ is \emph{visible} to $t_j$, denoted by $t_i \rightarrow t_j$, if and only if the writes of $t_i$ are all accessible to $t_j$ during the entire lifespan of $t_j$. We say that $t_i$ is \emph{invisible} to $t_j$, denoted by $t_i \nrightarrow t_j$, if and only if none of the writes of $t_i$ is accessible to $t_j$ during the entire lifespan of $t_j$.
\end{definition}

\begin{figure}
\centering
\includegraphics[width=.45\textwidth]{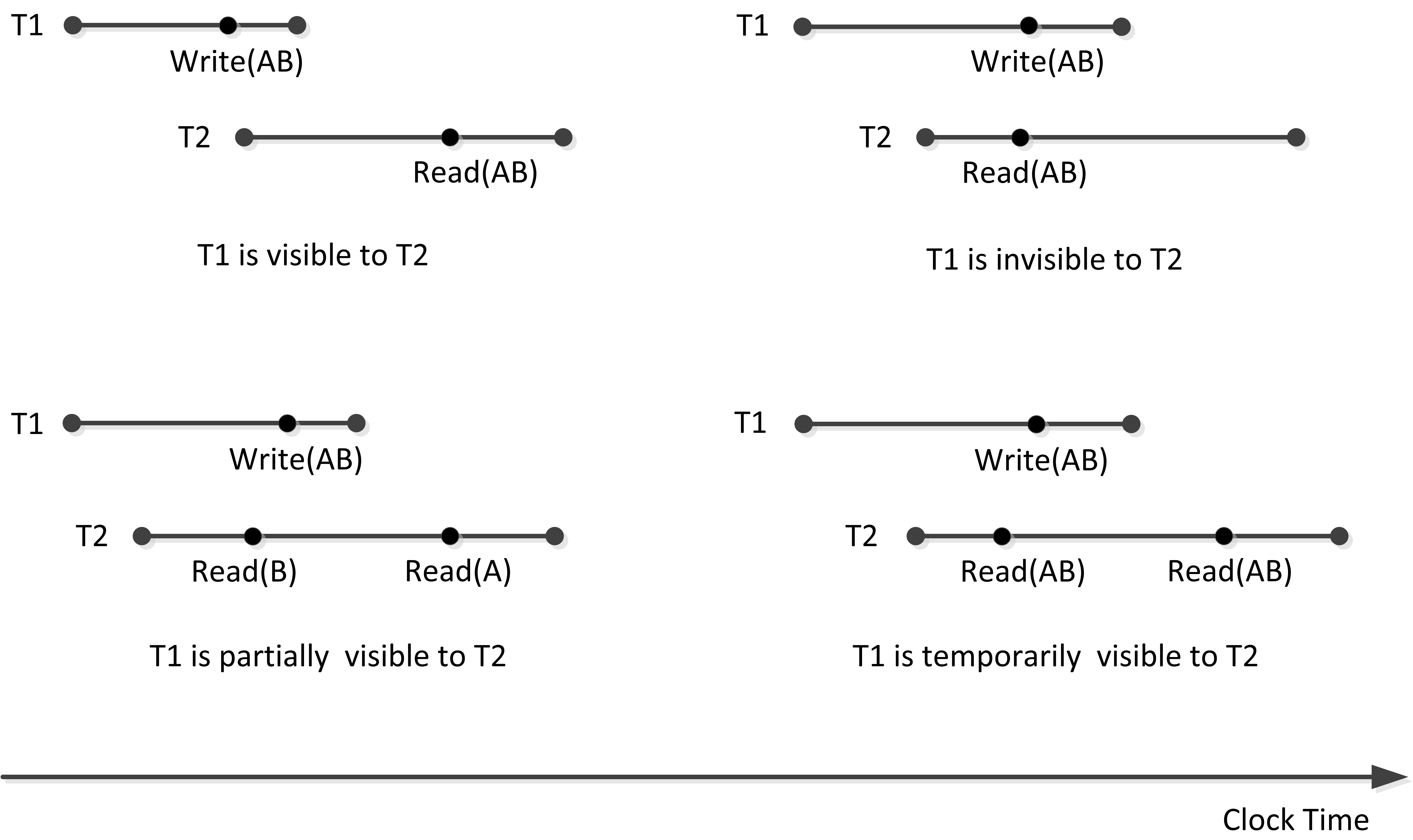}
\caption{Cases of Visibility
(A and B are the only data shared by T1 and T2.)}
\label{fig:visibility}
\end{figure}

In the definition, the writes of a transaction refer to committed writes. We assume that uncommitted writes (intermediate results) are internal data of a transaction and are thus invisible externally. Apart from the \emph{visible} and \emph{invisible} relationships, $t_i$ can be \emph{partially visible} to $t_j$, that is, only a subset of the writes of $t_i$ are accessible to $t_j$. Partial visibility is well possible in a history of transaction execution. However, as it is usually not desirable, we exclude it from our discussion. We do not consider \emph{temporary visibility} either. In other words, a transaction cannot be sometimes visible and sometimes invisible to another transaction. (See Figure~\ref{fig:visibility} for illustration.)

\begin{figure*}
\centering
\includegraphics[width=1.0\textwidth]{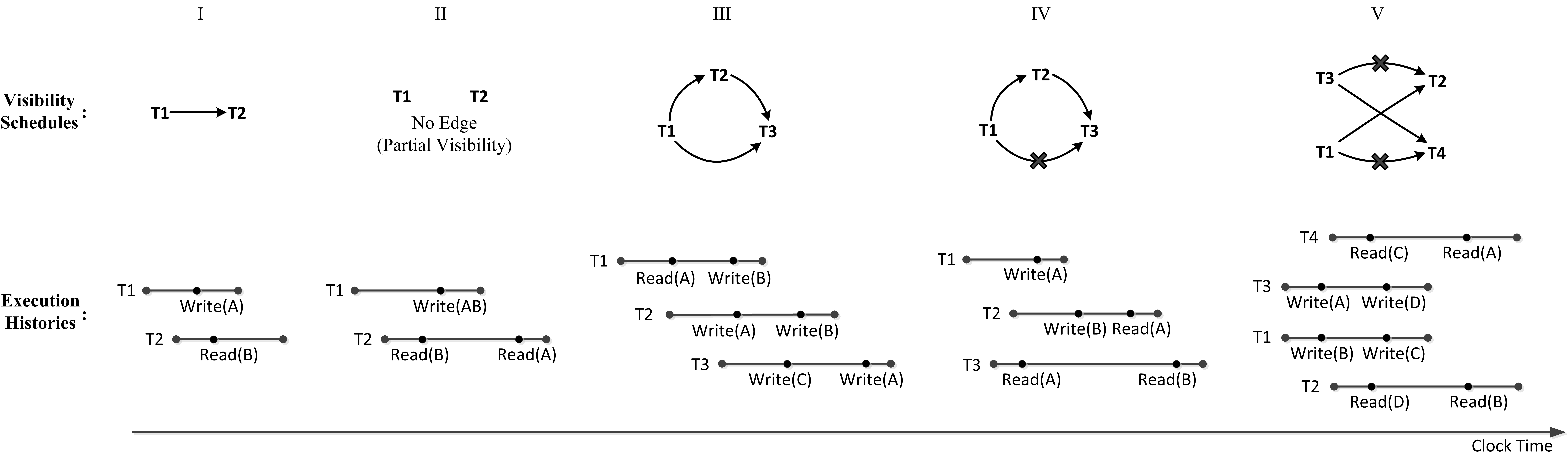}
\caption{Various Execution Histories and Their Visibility Schedules (Not all visibility relationship is instantiated in the schedules. For instance, in Schedule III, it is not difficult to infer $t_2 \nrightarrow t_1$, $t_3 \nrightarrow t_2$ and $t_1 \nrightarrow t_3$. In Schedule IV, we can have either $t_3 \rightarrow t_1$ or $t_3 \nrightarrow t_1$. For simplicity, we neglect the visibility relationships that are straightforward or unimportant.)}
\label{fig:schedules}
\vspace{0mm}
\end{figure*}

The concept of visibility is different from that of data dependency, by which traditional theories of concurrency control depict the relationship among transactions. Traditional approaches \cite{weikum2001transactional} usually assume a linear history of transaction processing, which is a flow of interleaved read and write operations issued by a set of transactions. This leads to three types of data dependency between a pair of transactions:
\begin{itemize}
  \item $wr$-dependency (a.k.a. flow dependency): if $t_i$ writes data $A$ before $t_j$ reads $A$, $t_j$ is $wr$-dependent on $t_i$, denoted by $t_i\xrightarrow{wr}t_j$; 
  \item $rw$-dependency (a.k.a. anti-Dependency): if $t_i$ reads data $A$ before $t_j$ writes $A$, $t_j$ is $rw$-dependent on $t_i$, denoted by $t_i\xrightarrow{rw}t_j$; 
  \item $ww$-dependency (a.k.a. output dependency): if $t_i$ writes data $A$ before $t_j$ writes $A$, $t_j$ is $ww$-dependent on $t_i$, denoted by $t_i\xrightarrow{ww}t_j$; 
\end{itemize}

As we can see, data dependency is event based. Only after a conflict occurs between two transactions, can a data dependency exist. In contrast, visibility is fact based. Even if there is no data dependency between two transactions, there can still be a visibility relationship among them -- while Transaction A does not access the data installed by Transaction B, it does not necessarily means that the data of B is not visible to A. Nevertheless, visibility relationship is constrained by data dependency. For instance, if $t_i\xrightarrow{wr}t_j$, we can conclude that $t_i \nrightarrow t_j$ is impossible, as $t_j$ has already accessed a write of $t_i$. Similarly, if $t_j\xrightarrow{rw}t_i$, we can conclude that $t_i \rightarrow t_j$ is impossible, as $t_j$ missed a write of $t_i$.

In summary, the definition of visibility relationship implies the following rules:
\begin{definition}
\emph{(Implication of Visibility)}: given a pair of transactions $t_i$ and $t_j$,
\begin{itemize}
\item[C1:] The visibility from $t_i$ to $t_j$ must fall in one and only one of the following cases: (i) $t_i \rightarrow t_j$; (ii) $t_i \nrightarrow t_j$; (iii) neither $t_i \rightarrow t_j$ nor $t_i \nrightarrow t_j$ (the case of partial or temporal visibility).
\item[C2:] If $t_i \rightarrow t_j$ , then $t_j \rightarrow t_i$ does not hold. (No mutual visibility.)
\item[C3:] If $t_j\xrightarrow{rw}t_i$, then $t_i \rightarrow t_j$ does not hold.
\item[C4:] If $t_i\xrightarrow{wr}t_j$, then $t_i \nrightarrow t_j$ does not hold.
\item[C5:] If $t_i\xrightarrow{ww}t_j$, then $t_i \nrightarrow t_j$ does not hold.
\item[C6:] If $t_i \rightarrow t_j$, then $t_j$ must not access the data versions previous to the ones installed by $t_i$.
\end{itemize}
\end{definition}
In theory, C2 is not mandatory. Mutual visibility is possible in a real history of transaction execution. However, it is of little help to the practice of concurrency control, and will only complicate our discussion. Therefore, we forbid mutual visibility in ViCC.

Given an execution history of a set of transactions, we can assign visibility relationship to the transactions based on the rules of Definition~2. If none of the rules is violated, we regard the assignment of visibility relationship valid. We call such an assignment a \emph{visibility schedule}.

\begin{definition}
\emph{(Visibility Schedule)}:
Given an execution history $X$ of a set of transactions $T = \{t_0, t_1, t_2, ..., t_n\}$, a \emph{visibility schedule} of $X$ is a function $S: T\times T \Rightarrow \{visible, invisible, \varnothing\}$, which obeys the rules in Definition~2.
\end{definition}

Figure~\ref{fig:schedules} illustrate some execution histories and their visibility schedules.
It is well possible that there are more than one visibility schedules for a particular execution history -- if there is no data dependency between two transactions, their visibility relationship can be arbitrary. For instance, in History I, although there is no dependency between $t_1$ and $t_2$, it is valid to set $t_1 \rightarrow t_2$ or $t_2 \rightarrow t_1$. In contrast, in History II, as $t_1$ is partially visible to $t_2$, neither $t_1 \rightarrow t_2$ nor $t_1 \nrightarrow t_2$ is valid.

\subsection{Isolation based on Visibility}


Based on the concept of visibility, we can define a number of isolation levels.

\subsubsection{Consistent Visibility}

The baseline among our isolation levels is called Consistent Visibility, which requires the visibility relationship between any pair of transactions to be \emph{atomic}. In other words, it does not allow partial or temporary visibility.

\begin{definition}
\emph{(Consistent Visibility (CV))}:
Let $X$ be an execution history of a set of transactions $T = \{t_0, t_1, t_2, ..., t_n\}$. We say that $X$ satisfies \emph{consistent visibility}, if and only if there is a visibility schedule of $X$ that satisfies the following criterion:
given any pair $t_i, t_j \in T (i\neq j)$, either $t_i \rightarrow t_j$ or $t_i \nrightarrow t_j$. (In other words, no partial or temporary visibility is allowed.)
\end{definition}


As we do not allow mutual visibility (Definition~2 C2),  $t_i \rightarrow t_j$ implies $t_j \nrightarrow t_i$ under CV.
Therefore, CV imposes an order between each pair of transactions, that is, if $t_i \rightarrow t_j$, then $t_i$ is regarded as transaction that commits prior to the start of $t_j$. However, this order is not a total order, as two transactions can be invisible to each other. As a result, there can only be three types of relationship between a pair of transactions -- $t_i \rightarrow t_j$ (implying $t_j \nrightarrow t_i$), $t_j \rightarrow t_i$ (implying $t_i \nrightarrow t_j$) or $t_i \nrightarrow t_j \wedge t_j \nrightarrow t_i$. Schedules III, IV and V in Figure~\ref{fig:schedules} are all CV schedules.



CV is an isolation level stronger than Read Committed and Repeatable Read (by definition), which suffer from partial or temporary visibility.
The name of atomic visibility has been mentioned in \cite{bailis2014scalable}, which proposed an isolation level called Read Atomicity (RA). However, since RA does not concern the order of write operations, it is strictly weaker than CV. As pointed out in Bailis' thesis \cite{bailis2015coordination}, RA tolerates the anomalies of Lost Updates and Missing Dependencies, as defined in Adya's thesis \cite{adya1999weak}. In contract, CV does not tolerate these anomalies. 

In the hierarchies of isolation levels described in Adya's thesis \cite{adya1999weak}, CV should be placed between Monotonic View (PL-2L) and Consistent View (PL-2+). CV is stronger than Monotonic View, which does not guarantee atomic visibility, and weaker than Consistent View, which requires visibility to be transitive (for instance, Schedule IV in Figure~\ref{fig:schedules} violates transitivity).

\subsubsection{Posterior Snapshot Isolation}

The second isolation level of ViCC is Snapshot Isolation (SI), which requires transactions to follow a temporal order. To avoid allocating timestamps at the beginning of each transaction, we redefine SI using the concept of visibility.

\begin{definition}
\emph{(Posterior Snapshot Isolation (PostSI))}:
Let $(s, c)$ denote a time interval, where $s$ and $c$ are two integers (representing the start time and commit time respectively) and $s < c$. Let $I$ denote the domain of time intervals.
Let $S$ be a CV schedule of a set of transactions $T = \{t_0, t_1, t_2, ..., t_n\}$. $S$ is also an \emph{SI} schedule, iff there is a function from $T$ to the domain of time intervals $I$, i.e., $F: T \Rightarrow I$, such that for any pair of transactions $t_i$ and $t_j$, their time intervals, $F(t_i) = (s_i, c_i)$ and $F(t_j) = (s_j, c_j)$, meet the following constraints:\\
(i) if $t_i \rightarrow t_j$, then $c_i \leq s_j$;\\
(ii) if $t_i \nrightarrow t_j$, then $c_i > s_j$.
\end{definition}

PostSI is a redefinition of SI. A PostSI schedule must first be a CV schedule. Besides, it requires that the visibility relationship among transactions abides by their temporal relationship -- a transaction sees all the transactions that commit prior to its start, and cannot see any transaction that commits after its start. In contrast to traditional definition of SI, PostSI does not rely on a real clock to define the temporal relationship among transactions. Instead, it determines the temporal relationship post priori based on the visibility relationship. As long as there exists a mapping from the visibility relationship to a linear timeline, SI can be satisfied.

\begin{figure}
\centering
\includegraphics[width=.38\textwidth]{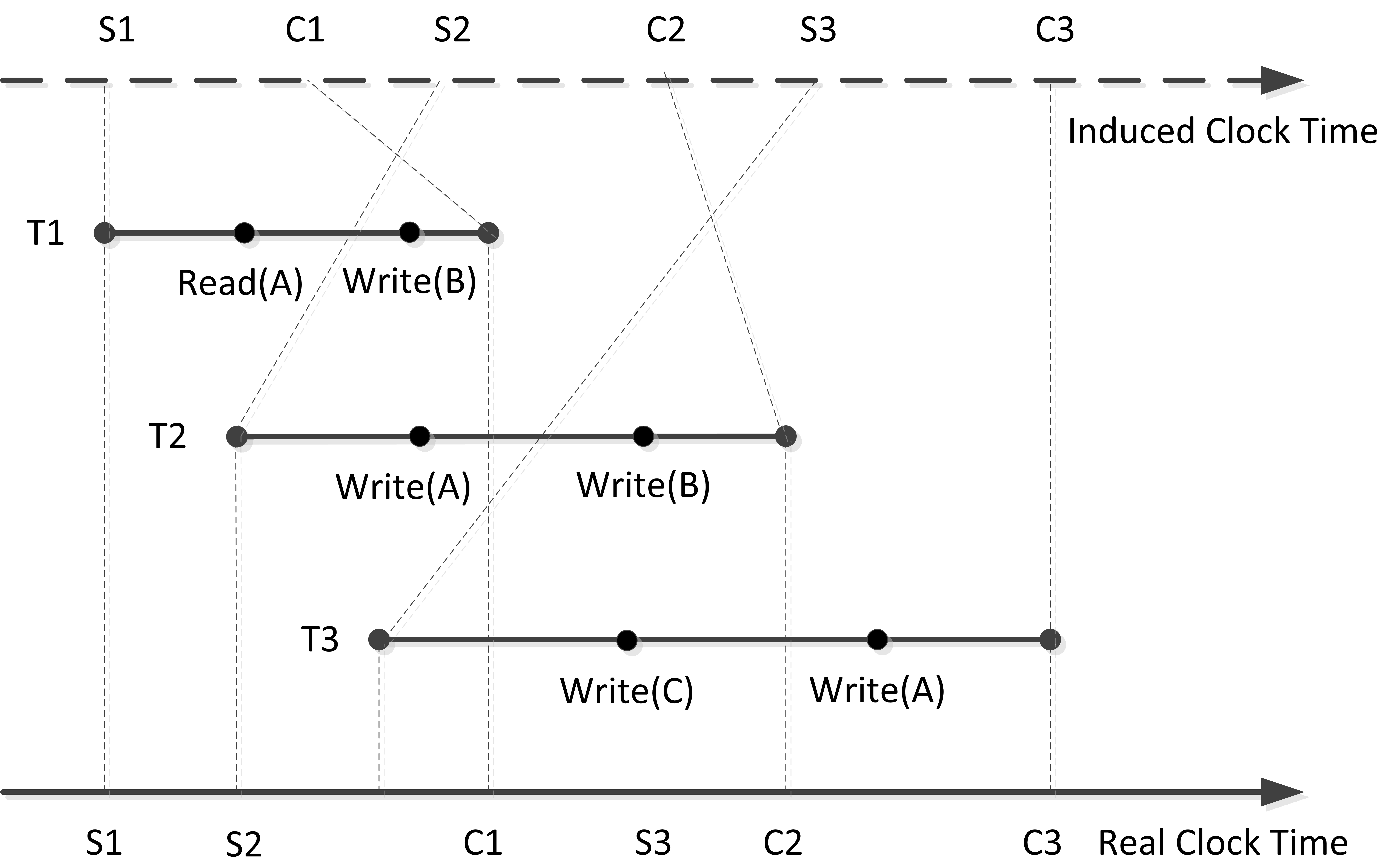}
\caption{Timestamp Assignment for Schedule III}
\label{fig:si}
\vspace{0mm}
\end{figure}

For instance, Schedule~III in Figure~\ref{fig:schedules} is a PostSI schedule, as we can find appropriate start and commit time for each of its transactions. The timeline at the top of Figure~\ref{fig:si} shows the start and commit times that can be induced from their visibility relationship. Based on the physical start and commit time of the transactions, this schedule is not valid to a traditional SI scheduler, which regards $t_1, t_2$ and $t_2, t_3$ as two pairs of conflicting transactions. However, it is a plausible schedule for PostSI, which uses logical timestamps. In contrast, Schedules IV and V in Figure~\ref{fig:schedules} are not PostSI schedules, as it is impossible to find appropriate start and commit time for their transactions.
If we consider only final effects, PostSI complies with the conventional definition of SI \cite{adya1999weak,ports2012serializable}.

There have been several variants of SI's definition in the literature. The Generalized SI defined in \cite{elnikety2005database} shares some spirit of our definition of PostSI. It allows a transaction to set an earlier start time than its actual start. In this sense, our definition is even more general or relaxed, as we allow both the start time and commit time to deviate from the actual start and end of a transaction. Time is only logical to PostSI.

PostSI poses stronger requirements to schedules than CV. There are schedules that satisfy the criteria of CV but violate that of SI. Schedules IV and V in Figure~\ref{fig:schedules} are two examples.
As Schedule IV shows, CV does not satisfy transitivity, i.e., $t_1 \rightarrow t_2$, $t_2 \rightarrow t_3$ and $t_1 \nrightarrow t_3$, while SI requires visibility to be transitive.
In Schedule V, based on the transactions' operations on $A,B,C,D$, we have $t_1 \rightarrow t_2$, $t_3\rightarrow t_4$, $t_3 \nrightarrow t_2$, and $t_1\nrightarrow t_4$. To meet the criteria of SI, if $t_1 \rightarrow t_2$, then $t_1$'s logic commit time must be earlier than $t_2$'s logic start time, i.e., $c_1 < s_2$. Analogously, we can deduce $c_3 < s_4$ from $t_3\rightarrow t_4$, $s_2<c_3$ from $t_3 \nrightarrow t_2$, and $s_4<c_1$ from $t_1 \nrightarrow t_4$. The inequations $c_1 < s_2$, $s_2<c_3$, $c_3 < s_4$ and $s_4<c_1$ are cyclic and cannot be all satisfied. Therefore, Schedule~V is not a SI schedule either.



Definition~5 further leads us to the following law, which provides a clearer picture about the difference between CV and PostSI.

\newtheorem{theorem}{Theorem}
\begin{theorem}
Let $S$ be a CV schedule of a set of transactions $T = \{t_0, t_1, t_2, ..., t_n\}$.
Suppose $\prec$ is an order of $T$, such that
\begin{itemize}
\item $t_i \prec t_j$, iff $t_i \rightarrow t_j$;
\item $t_j \preccurlyeq t_i$, iff $t_i \nrightarrow t_j$;
\end{itemize}
$S$ is an PostSI schedule, if and only if: if $\prec$ contains a cycle, then the cycle must contain two consecutive edges of invisibility in the form of $t_i \nrightarrow t_k \nrightarrow t_j$ (or $t_j \preccurlyeq t_k \preccurlyeq t_i$).
\end{theorem}

Theorem~1 states that a CV schedule is an PostSI schedule, if its order $\prec$ is either acyclic, or each cycle in $\prec$ comprises two consecutive invisibility relationships. As illustrated by Schedules IV and V in Figure~\ref{fig:schedules}, the schedules do not satisfy PostSI, just because their $\prec$ orders are cyclic (i.e., $t_1 \prec t_2 \prec t_3 \preccurlyeq t_1$ and $t_1 \prec t_2 \preccurlyeq t_3 \prec t_4 \preccurlyeq t_1$), and the cycles do not contain consecutive invisibility. The proof of Theorem~2 can be analogized to that of the theories behind Serializable SI~\cite{fekete2005making}.

\subsubsection{Serializable Visibility}

Similar to SI, we can define Serializability on top of CV. We name it Serializable Visibility.

\begin{definition}
\emph{(Serializable Visibility (SV))}:
Let $S$ be a CV schedule of a set of transactions $T = \{t_0, t_1, t_2, ..., t_n\}$.
$S$ satisfies \emph{serializable visibility}, if and only if there exist a visibility relationship among $T$ such that:
(i) For each pair of transactions $t_i$ and $t_j$, if $t_i \nrightarrow t_j$, then $t_j \rightarrow t_i$;
(ii) The $\rightarrow$ relationship is acyclic;
\end{definition}

\begin{theorem}
A scheduler satisfying serializable visibility is serializable.
\end{theorem}

Serializability requires a total order of transactions, such that each transaction is visible to the transactions behind it. In contrast, CV and PostSI allow two transactions to be mutually invisible.

The definitions of CV, PostSI and SV are all based on visibility. This allows us to implement a single concurrency controller that supports all the three isolation levels.

\section{Scheduling Methods}

In what follows, we introduce a scheduler for CV and then extend it to PostSI and SV. To ensure that a scheduler satisfies an isolation level defined in the previous section, the system should only permit the visibility relationship allowed by that isolation level.

\subsection{A Scheduler for CV }

As the definition of CV does not involve the concept of clock, it is relatively easy to come up with a CV scheduler that needs no centralized coordination.
The main concern of CV is atomic visibility. According to Definition~3, the main task of a CV scheduler is to ensure that the data dependencies among transactions does not violate atomic visibility.

\begin{figure*}
\centering
\includegraphics[width=.92\textwidth]{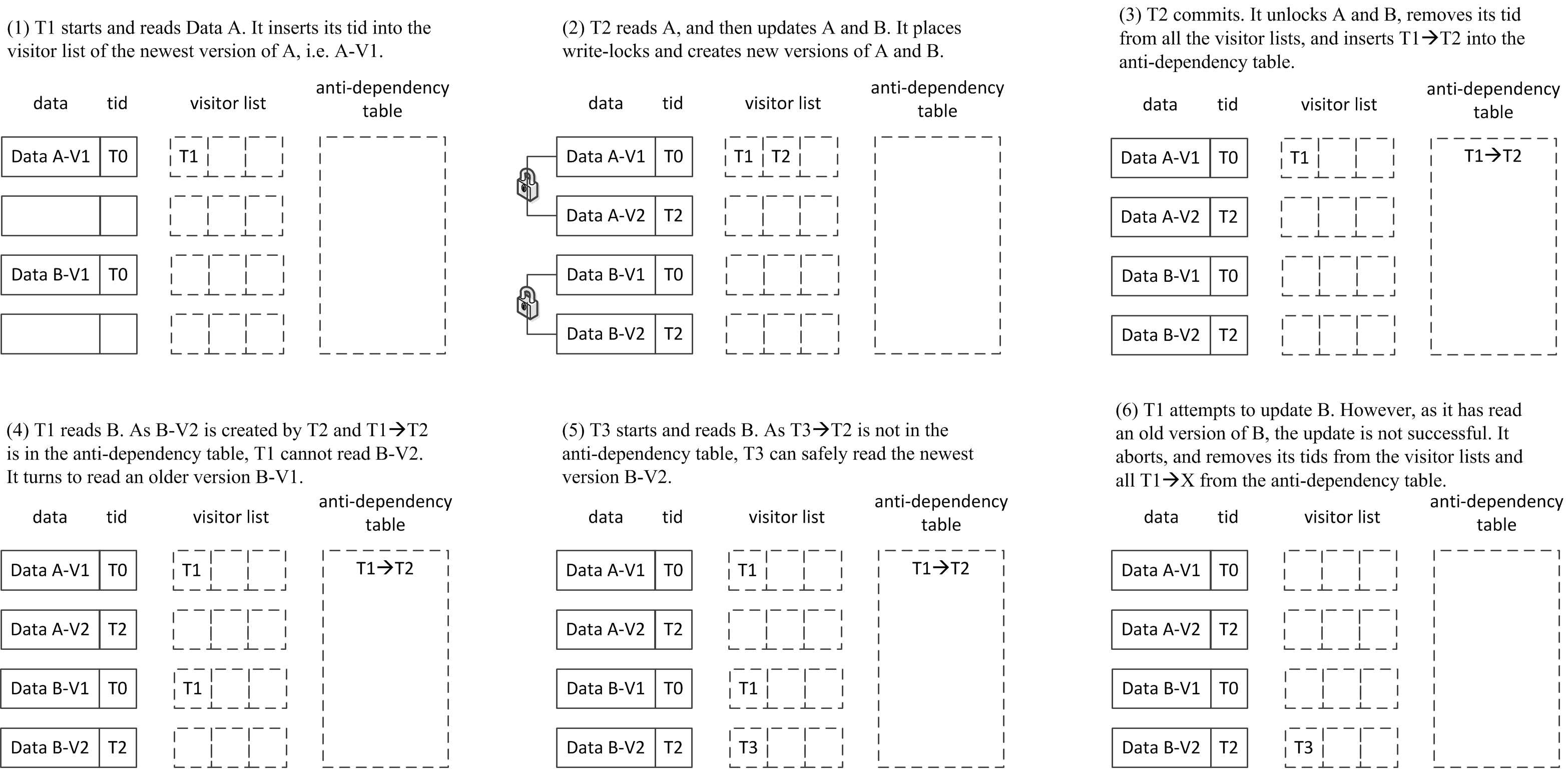}
\caption{An Example of Transaction Execution Procedure that Follows CV}
\label{fig:cv-example}
\vspace{0mm}
\end{figure*}

First, if there is a $rw$ dependency between $t_i$ and $t_j$, i.e., $t_i\xrightarrow{rw}t_j$, we must ensure that $t_j$ is invisible to $t_i$, i.e., $t_j \nrightarrow t_i$ (Definition~3 (i)). In other words, if $t_j$ overwrites a data version $t_i$ has read, $t_j$ must be invisible to $t_i$ and $t_i$ must not read any data generated by $t_j$. This can be achieved by keeping track of all the $rw$ dependencies among the ongoing transactions -- whenever $t_i$ attempts to read a data version generated by $t_j$, it first checks whether $t_i\xrightarrow{rw}t_j$ exists: if $t_i\xrightarrow{rw}t_j$ does not exist, $t_i$ proceeds to read the data; if it exists, $t_i$ turns to try an older version of the data.

Second, if there is a $wr$ dependency between $t_i$ and $t_j$, i.e., $t_i\xrightarrow{wr}t_j$, we must ensure $t_i\rightarrow t_j$ (Definition~3 (ii)). In other words, if $t_j$ reads a data version generated by $t_i$, then $t_i$ is visible to $t_j$ and all the other data generated by $t_i$ should be visible to $t_j$ too. This can be violated, if we let a transaction always read the freshest versions of the data. For instance, if $t_i$ commits during the execution of $t_j$, it is possible that $t_j$ happens to read an old data version overwritten by $t_i$ and a new data version created by $t_i$. However, if the system can keep track of the $rw$ dependencies, this case can be avoided --
when $t_i$ commits, it should check if it has overwritten any data version read by $t_j$; if it has, $t_j\xrightarrow{rw}t_i$ is recorded; then, $t_j$ is forbidden to read any data generated by $t_i$.

As we can see, atomic visibility can be achieved by tracing the anti-dependency among the ongoing transactions. Thus, our CV scheduler works as follows.

\vspace{3mm}
\noindent{\bf A CV Scheduler:}
\begin{enumerate}[(1)]
  \item Each transaction is assigned an unique TID. (To generate unique TIDs in parallel, each session can create a TID as a concatenation of its session id and an id from its local counter.)
  \item Each version of a data item is associated with a TID, recording the transaction that created this version. Each version also corresponds to a \emph{visitor list}, which records the TIDs of the ongoing transactions that has read this version.
  \item The scheduler maintains an anti-dependency table that records the $rw$-dependency between the ongoing transactions. Each entry in the table is in the form of $t_j\xrightarrow{rw}t_i$.
  \item When a transaction $t_j$ reads a data item, it always reads the latest version that is visible. It starts with the latest version. If the version's TID is $t_i$ and it can find $t_j\xrightarrow{rw}t_i$ in the anti-dependency table, $t_j$ regards this version invisible (because $t_j\xrightarrow{rw}t_i$ implies $t_i\nrightarrow t_j$). Then, $t_j$ has to try the second latest version and so on. Once $t_j$ finds the latest visible version, it performs the read, and during the read operation it adds its TID to the visitor list of the version.
  \item When a transaction $t_j$ writes to a data item, it places a write lock on the item. It unlocks the item only when it commits. Upon the commit, the data versions created by $t_j$ is immediately visible to other transactions. After $t_j$ obtains the write lock, it checks if the following criteria can be satisfied: (i) if $t_j$ has read the data item, the version it has read must be the newest version; (ii) if the newest version's TID is $t_i$, then $t_j\xrightarrow{rw}t_i$ must not be in the anti-dependency table. If one of the two criteria is violated, $t_j$ has to abort, because one of its concurrent transactions has updated the version.
  \item When a transaction $t_j$ commits, for every TID in the visitor list of each data version it has updated (let the TID be $t_i$), it adds $t_i\xrightarrow{rw}t_j$ to the anti-dependency table. After the commit, we safely remove $t_j$ from all the visitor lists, and every $t_j\xrightarrow{rw}t_k$ from the anti-dependency table.
\end{enumerate}

The examples in Figure~\ref{fig:cv-example} illustrate how the CV scheduler executes transactions. The instruments of visitor lists and the anti-dependency table provide sufficient information to determine the visibility relationship between transactions. We can thus ensure that a transaction is either visible or invisible to another transaction. Rule (5) disallows two concurrent transaction to modify the same piece of data. Thus, the scheduler ensures a total order of the write operations. According to Definition~5, this scheduler can enforce CV. Most importantly, the CV scheduler can be completely decentralized. It does not require any central data structure for coordination, as the visitor lists and the anti-dependency table can all be distributed.

Despite the fact that CV is weaker than SI and serializablity in isolation, it can be a practical solution to applications that do not require strong data consistency. Due to lack of space, we schedule the elaboration of CV's use cases in our future work.

\subsection{A Scheduler for PostSI}






Our scheduler of PostSI is built on top of the CV scheduler.
According to Definition~5, if a scheduler can assign an appropriate time interval to each transaction, it can ensure that the resulting schedule is snapshot isolated.
To avoid centralized coordination, our SI scheduler does not rely on a central clock to determine the time intervals.
It leaves to the transactions to decide their own start and commit time through negotiation.
For each transaction, our PostSI scheduler maintains a lower and an upper bounds of its start time, i.e., $[\underline{s},\overline{s}]$, and a lower bound of its commit time, i.e., $[\underline{c},+\infty)$. (We do not consider the upper bound of commit time, as it can technically be arbitrarily large.) Initially, $\underline{s}=0$, $\overline{s}=+\infty$ and $\underline{c}=0$. During the execution, we adjust the $\underline{s},\overline{s},\underline{c}$ according to the visibility relationships between the transaction and the others. At the end of the transaction, we pick a valid start and a valid commit time based on their lower and upper bounds. If no start or commit time is valid, the transaction has to abort.

Therefore, our PostSI scheduler can be realized by complementing the CV scheduler with the following additional rules.

\vspace{3mm}
\noindent{\bf Complementary Rules of A PostSI Scheduler:}
\begin{enumerate}[(1)]
  \item When a transaction $t_j$ starts, the lower and upper bounds of its start time and the lower bound of its commit time are initialized as $\underline{s_j}=0$, $\overline{s_j}=+\infty$ and $\underline{c_j}=0$.

  \item Each version of a data item is associated with two timestamps, CID and SID. CID records the commit time of the transaction that created this version. The SID records the maximum start time of the transactions that have read this version.

  \item If a transaction $t_j$ reads or overwrites a data version, then the transaction that created the version should be visible to $t_j$. If the data version's CID is $cid$,
        $t_j$ updates the lower bounds of its start and commit time, by setting $\underline{s_j}= max(\{\underline{s_j},cid\})$ and $\underline{c_j}= max(\{\underline{c_j},cid\})$.

  \item When a transaction $t_j$ commits, it performs the following actions:
      \begin{enumerate}
          \item {Determining its own time interval:}
           $t_j$ sets its start time to $s_j = \underline{s_j}$.
           Let $S$ be the SIDs of the data versions $t_j$ has read. We set $\underline{c_j}= max(\{\underline{c_j}\}\cup S)$.
           For every $t_i$ such that $t_i\xrightarrow{rw}t_j$ can be found in the anti-dependency table, we set $\underline{c_j}= max(\{\underline{c_j},\underline{s_i}\})$. Finally, we set the commit time of $t_j$ to $c_j = max(\{\underline{c_j},s_j\})+1$.
          \item {Adjusting the orders of its conflicting transactions:} For every ongoing $t_k$ such that $t_j\xrightarrow{rw}t_k$, we set $\underline{c_k} = max(\{\underline{c_k}, s_j+1\})$ (as $t_j\xrightarrow{rw}t_k$ implies $t_k \nrightarrow t_j$ and $c_k>s_j$).
          For every ongoing $t_i$ such that $t_i\xrightarrow{rw}t_j$, we set $\overline{s_i} = min(\{\overline{s_i}, c_j-1\})$ (as $t_i\xrightarrow{rw}t_j$ implies $t_j \nrightarrow t_i$ and $c_j>s_i$).
          \item {Setting SIDs and CIDs:}  we sets the CIDs of the data created by $t_j$ to $c_j$. For each data version $t_j$ has read, if its SID is smaller than $s_j$, we set its SID to $s_j$.
      \end {enumerate}
  \item At anytime during the execution of a transaction $t_j$, if $\underline{s_j} > \overline{s_j}$, $t_j$ has to abort, as it is no longer possible to find a valid start time for $t_j$.
\end{enumerate}

Basically, when we encounter the visibility  $t_i \rightarrow t_j$, we raise the lower bound of $t_j$'s start time to at least the commit time of $t_i$. This is enforced by Rule~(3). When encountering the visibility $t_i \nrightarrow t_j$, we either lower the upper bound of $t_j$'s start time to be smaller than the commit time of $t_i$, or raise the lower bound of $t_i$'s commit time to be greater than the start time of $t_j$, depending on whether $t_i$ or $t_j$ commits first. This is enforced by Rule~(4). The anti-dependency table records only the $rw$ dependencies of the ongoing transactions. If $t_j$ commits before $t_i$ starts, $t_j\xrightarrow{rw}t_i$ (i.e., $t_i \nrightarrow t_j$) will not be found in the anti-dependency table. In this case, the scheduler passes $t_j$'s start time to $t_i$ using SIDs. This is the only mission of SIDs.

When a transaction finalizes its time interval (Rule~(4)(a)), it needs to ensure that the upper and lower bounds are not violated, i.e., $\underline{s_j} \leq s_j \leq \overline{s_j}$ and $\underline{c_j} \leq c_j$. Deciding the start time $s_j$ is simple, as $s_j$ will not interfere with the future transactions -- our scheduler simply sets $s_j=\underline{s_j}$. Deciding the commit time $c_j$ requires more thought. On the one hand, $c_j$ may lower the $\overline{s}$ of the transactions that regard $t_j$ invisible (Rule~(4)(b)).
On the other hand, it will also be used as the CIDs of the updated data (Rule~(4)(c)), which will in turn raise the $\underline{s}$ of the future transactions reading the data.
If $c_j$ is too small or too large, it may cause other transactions to abort.
To minimize the chance of abort, our scheduler sets $c_j$ to the smallest value that is larger than the $\underline{s}$ of the transactions regarding $t_j$ invisible (as stated in Rule~(4)(a)).

If we apply the PostSI scheduler to Schedules IV and V in Figure~\ref{fig:schedules}, they will not be allowed to pass. $t_3$ of Schedule IV will not be allowed to read the newest version of $B$, as it will make $\underline{s_3}$ greater than $\overline{s_3}$. Similarly, neither $t_4$ nor $t_2$ of Schedule V is allowed to proceed when attempting to read the newest versions of $A$ and $B$.

According to Rule~(4), when $t_j$ commits, for every $t_i\xrightarrow{rw}t_j$ and $t_j\xrightarrow{rw}t_k$, it should inform $t_i$ and $t_k$ about its time interval. This is where the concurrent transactions conduct negotiation. However, $t_i$ or $t_k$ may fail to receive the message of $t_j$, if they commit before the message arrives. Nevertheless, when $t_i$ and $t_k$ commit, they will initiatively send their orders to $t_j$. Thus, we can guarantee that the message from at least one direction will arrive safely. The negotiation is guaranteed to take place.

\subsection{A Scheduler for SV}

There are efficient methods to upgrade an SI scheduler to a serializability scheduler~\cite{fekete2005making}. They can be applied to our case. However, it is simpler to directly build the SV scheduler on top of our CV scheduler. According to Definition~6, when atomic visibility is already guaranteed, we only need to impose a total order to the visibility relationship. Similar to what is done in the PostSI, our SV scheduler lets transactions negotiate about their own orders.

The SV scheduler is realized by complementing the CV scheduler with the following rules.

\vspace{3mm}
\noindent{\bf Complementary Rules of An SV Scheduler:}
\begin{enumerate}[(1)]
  \item When a transaction $t_j$ starts, the lower and upper bounds of its order $o_j$ are initialized as $\underline{o_j}=0$ and $\overline{o_j}=+\infty$.

  \item Each version of a data item is associated with two timestamps, CID and SID. CID records the order of the transaction that created this version. SID records the maximum order of the transactions that have read this version.

  \item If a transaction $t_j$ reads or overwrites a data version, then the transaction that created the version should be visible to $t_j$. If the data version's CID is $cid$,
        $t_j$ updates the lower bounds of its order by setting $\underline{o_j}= max(\{\underline{o_j},cid\})$.

  \item When a transaction $t_j$ commits, it performs the following actions:
      \begin{enumerate}
          \item {Determining the order of itself:}
           Let $S$ be the SIDs of the data versions $t_j$ has read. We set $\underline{o_j}= max(\{\underline{o_j}\}\cup S)$.
           For every $t_i$ such that $t_i\xrightarrow{rw}t_j$ can be found in the anti-dependency table, we set $\underline{o_j}= max(\{\underline{o_j},\underline{o_i}\})$. Finally, we set the order of $t_j$ to $o_j = \underline{o_j}$.
          \item {Adjusting the orders of its conflicting transactions:} For every ongoing $t_k$ such that $t_j\xrightarrow{rw}t_k$, we set $\underline{o_k} = max(\{\underline{o_k}, o_j+1\})$ (as $t_j\xrightarrow{rw}t_k$ implies $t_k \nrightarrow t_j$ and $o_k>o_j$).
          For every ongoing $t_i$ such that $t_i\xrightarrow{rw}t_j$, we set $\overline{o_i} = min(\{\overline{o_i}, =o_j-1\})$ (as $t_i\xrightarrow{rw}t_j$ implies $t_j \nrightarrow t_i$ and $o_j>o_i$).
          \item {Setting SIDs and CIDs:}  we sets the CIDs of the data created by $t_j$ to $o_j$. For each data version $t_j$ has read, if its SID is smaller than $o_j$, we set its SID to $o_j$.
      \end {enumerate}
  \item At anytime during the execution of a transaction $t_j$, if $\underline{o_j} > \overline{o_j}$, $t_j$ has to abort, as it is no longer possible to find a valid order for $t_j$.
\end{enumerate}

The SV scheduler follows the same procedure as that of the PostSI scheduler. It only needs to determine a single order for each transactions, while the PostSI scheduler needs to determine a time interval for each transaction, represented by the start time and the commit time.

In the above schedulers, we only considered read and write operations. In practice, deletion and insertion should also be taken care of to ensure the correctness of concurrency control. While phantom read is not a threat to MVCC, other anomalies are possible, as we may miss some data dependency if the schedulers ignore deletion and insertion. Fortunately, conventional solutions, such as predicate locking, are compatible with ViCC, and can be adopted to safeguard insertion and deletion. Careful engineering is required to make them efficient. In our current implementation of the schedulers, we do not consider insertion and deletion.

\section{The Implementation}


We have introduced three schedulers of ViCC that enforce different levels of isolation. As the PostSI and SV scheduler both build upon the CV scheduler, it is convenient to implement the three schedulers together in a single system, so that the system offers multiple degrees of isolation for software engineers to choose from.

Compared to traditional schedulers relying on central timestamp allocation, the ViCC schedulers introduce additional overheads to concurrency control.
While these overheads can be regarded as the price for eliminating centralized coordination, they can be minimized by careful engineering.
In what follows, we discuss how to implement the ViCC schedulers in an MPP database system with a shared-nothing architecture. To process transactions on a shared-nothing architecture, each transaction is allocated to a single computing node, known as the host of the transaction. The host will, in turn, distribute the work to other nodes, which work concurrently to accomplish the tasks of the transaction. Finally, the host employs a commit protocol, such as two-phase commit, to finish the transaction. Our implementation attempts to minimize the cost incurred by blocking and communication.



\subsection{Distribution of the Operational Data}

To implement a distributed ViCC scheduler, a critical issue is the management of the operational data, i.e., the visitor lists, the anti-dependency table and the bounds of each transaction's order. As the operational data is shared among different transactions, cross-node communication cannot be completely avoided. As such, we need to minimize the communication cost.

Data items and their visitor lists can certainly be collocated, as they are always accessed together. Then, no extra cross-node communication will be incurred. As we do not require visitor lists to be persistent, they can be detached from the data and stored in the memory. This can make their maintenance less costly. It is unlikely that all the data are accessed concurrently. Therefore, the space consumption of visitor lists is usually much smaller than that of the data.

For each $t_i\xrightarrow{rw}t_j$ in the anti-dependency table, we store it on both the node hosting $t_i$ and the node hosting $t_j$. Hence, insertion and deletion of the anti-dependencies requires cross-node communication, while lookup of the anti-dependency table can be performed locally most of the time.
We implement each anti-dependency table as a hash table, to facilitate its lookup.

The bounds of each transaction's order, which are required by PostSI and SV respectively, is maintained by the host of the transaction.
As stated in the PostSI scheduler (Rule~(3)), the lower bound of the start time (i.e., $\underline{s}$) needs to be updated upon each data access. When a transaction needs to process the data on a remote node, it delegates the work to the remote node. It sends its TID and a copy of its $\underline{s}$ to the remote node too, which can update the $\underline{s}$ locally while processing the data. After the remote node finishes its work, it sends its local $\underline{s}$ along with the results back to the host, which can update the global $\underline{s}$. The SV scheduler can do the same when maintaining $\underline{o}$.
When a transaction is about to commit, it needs to update the bounds of its conflicting transactions (Rule~(4)(b) of the PostSI and SV schedulers). This may incur cross-node communication.

\subsection{Optimizing Read Intensive Transactions}

A major advantage of MVCC lies in that it eliminates the blocking caused by read-write conflicts. This is especially beneficial to read intensive transactions, which may involve the execution of OLAP queries. ViCC is supposed to preserve this advantage of MVCC. Fortunately, read operations in ViCC are indeed nonblocking. Although each read operation implies an insertion in a visitor list, such an insertion can be completed within the same atomic operation of the read, without involving locking. Beside nonblocking read, other tactics need to be applied to ensure the efficiency of read intensive queries.

First, when a transaction performs a read, the CV scheduler requires it to lookup the anti-dependency table to confirm the data's visibility. If the read occurs on a remote read, the lookup will incur cross-node communication. This can be costly, especially for an OLAP-style query that needs to access a large amount of remote data. Fortunately, PostSI and SV do not need to lookup the anti-dependency table when reading data. Instead of using the anti-dependency table, a transaction can use CID to determine visibility -- a data item is visible, only if its CID is smaller than the upper bound of the transaction's start time. Thus, the CV scheduler can be optimized by employing CIDs too.

Second, when performing parallel query processing under PostSI and SV, a transaction distributes the bounds of its time interval and order (i.e., $\underline{s}$ and $\underline{o}$) to the remote nodes, so that they could update the bounds locally. At the same time, the bounds (i.e., $\overline{s}$ or $\overline{o}$) on the host can possibly be updated by a conflicting transaction (Rule~(4)(b) of PostSI and SV). When the transaction receives the bounds back from the remote nodes, it may find that the bounds are violated (i.e., $\underline{s} > \overline{s}$ or $\underline{o} > \overline{o}$) and have to abort. The same abort can occur repeatedly, if the transaction happens to read a hot remote item that is frequently updated. As a remedy, when a transaction aborts, we can retry the transaction by fixing the initial upper bounds (i.e., $\overline{s}$ or $\overline{o}$) at the highest CID the transaction encounters before its previous abort. During the retry, the transaction can avoid accessing the newest data whose CID is higher than the upper bounds. Then, the same abort will not be repeated.

Third, when a transaction ends, it is supposed to update the SIDs of the data it has read and remove its TID from all the visitor lists. For a read intensive transaction, this can be costly. On the one hand, it requires a transaction to maintain a huge read set that memorizes all the data items it has read. On the other hand, it requires a transaction to perform a large amount of work in the commit phase, which may impair the performance of distributed transactions severely.

As a workaround, we can apply lazy deletion to visitor lists. Namely, when a transaction ends, it does not update the visitor lists immediately. Only when the next transaction accesses a visitor list will it remove the outdated TIDs from the list. TIDs are usually generated by incrementing a set of TID counters. The current values of the TID counters can be periodically broadcasted to the nodes, to enable the detection of outdated TIDs.

SID is meant to inform a transaction $t$ about the max start time or the max order of the committed transactions that regard $t$ invisible, so that $t$ can set its own commit time or order correctly.
If the deletion of the visitor lists is delayed, the update of SIDs can be delayed too, as the visitor lists reserve the information about invisibility. Specifically, when removing a transaction's TID from a visitor list, we check the start time or order of the transaction, and use it to update the SID of the data. This requires each computer node to keep the time interval or order of a committed transaction in its cache for a certain period.


With the optimization, a transaction does not need to maintain a read set. Neither does it need to update a large number of visitor lists and SIDs when committing. As a result, its commit phase can be significantly shortened. This gives ViCC an advantage over some OCC styled approaches, such as TicToc \cite{yu2016tictoc}, which have to perform intensive validation work in the commit phase, during which the data in the write set has to be locked.
ViCC does not require such a lengthy validation phase, because the use of anti-dependency table enables it to track the modification of its read set. The price ViCC pays for this advantage is the overhead of maintaining the anti-dependency table.

\subsection{The Commit Phase}


Similar to some OCC approaches \cite{tu2013speedy}, ViCC does not need to perform real write before the commit phase. Instead, each transaction can keep its write set private. Only when the transaction is about to commit, it applies its write set to the real data. This provides several benefits. First, the period of each write lock can be shortened, so as to enable higher concurrency. Second, deadlock can be avoided, as the write locks can be added in a strict order.

As shown in the PostSI and SV schedulers, three rounds of communication need to be conducted in the commit phase. In the first round, the write operations are materialized on the corresponding nodes and the data being updated is locked; at the same time, to determine the commit time of the transaction, the bounds of the time intervals and orders of its conflicting transactions are retrieved (Rule (4)(a) of PostSI and SV). In the second round, after the commit time is determined, the transaction needs to contact the conflicting transactions again to update their bounds of time intervals or orders; at the same time, the anti-dependency table needs to be updated. In the final round, the SIDs and CIDs of the data are updated, and the updated data is unlock.
To work with the two-phase commit protocol (2PC), the first round of communication can be integrated into the \emph{prepare} phase of 2PC, and the last round of communication can be integrated into the \emph{commit} phase of 2PC.  The second round of communication is additional to 2PC. Fortunately, this extra round of communication occurs only when there is contention. It can be avoided if a transaction does not conflict with the others.

There is a short interval after a transaction commits and before it makes its updated version available to other transactions. If another transaction happens to read its updated version during the interval, a race condition may arise. In our implementation, we maintain a writer list for each data item being updated. When a transaction, say $t_1$, updates a data item, it adds its TID to the writer list of the item. Only after $t_1$ commits, it removes its TID from the writer list. When another transaction $t_2$ reads the data item, it will first check the writer list. If $t_2$ finds $t_1$ in the writer list, it will regards $t_1$'s writes invisible (by enforcing $s_2<c_1$ in PostSI and $o_2<o_1$ in SV). This allows us to prevent the race condition during the commit phase.

\section{Experimental Evaluation}

In this section, we evaluate the performance of ViCC. We implemented the CV, PostSI and SV schedulers on a distributed in-memory key-value store, which we created for experiment's purpose. The KV-store supports point queries, simple range queries (using secondary hash indexes), as well as data manipulation operations such as insertion, deletion and update. When deployed on a cluster, the KV-Store partitions a database based on the key values, and assigns one partition to each node. Each node can host multiple sessions for processing user requests and multiple workers for processing transactions. A master node is responsible for maintaining the database, while it never participates in transaction processing.

\subsection{Experimental Setup}


Our experiments were conducted on a cluster of 30 virtual machines. Each virtual machine is equipped with two 4-core Intel Xeon E5620 2.4GHz processors and 32GB of DRAM. Each core has a private 32KB L1 cache and a private 256KB L2 cache. At the physical level, every 4 cores share a 12MB L3 cache.  The hypervisor of the virtual machines is the OpenStack released by Icehouse in April 2014. The operating systems running on the virtual machines are all CentOS 6.5. InfiniBand is used for networking. According our test, the bandwidth for end-to-end communication is around 1Gbps.

We compared our approaches against several SI schedulers in the literature. We chose SI schedulers because SI is one of the most efficient MVCC strategies that are widely adopted by real world systems. Moreover, there are several decentralized SI schedulers \cite{binnig2014distributed,du2013clock} that seem competitive to our approaches. We implemented the following SI schedulers in our systems:
\begin{enumerate}
\item We implemented a \emph{conventional SI scheduler}, by adopting the implementation of PostgreSQL 9.4. The SI scheduler of PostgreSQL uses a single timestamp and a snapshot of ongoing transactions to determine the time interval of each transaction. The timestamp represents the start time of the transaction. The snapshot contains the TIDs of all the ongoing transactions at the start time. 
    A central coordinator, residing on the master, is responsible for allocating timestamps and maintaining a snapshot of the ongoing TIDs. When a transaction starts, it obtains a timestamp and a snapshot from the coordinator. When a transaction terminates, it contacts with the coordinator again to remove its TID from the snapshot.
\item We implemented a \emph{optimal scheduler}, aiming to find out the best possible performance of SI. This scheduler follows the exact procedure of the conventional SI scheduler, except that it requires no centralized coordination. To eliminate the coordination, we simply assign an arbitrary timestamp and an empty snapshot to each transaction. Therefore, the optimal scheduler does not ensure correctness. It is intended to represent an upper bound of SI's performance.
\item We also implemented \emph{DSI} (the incremental snapshot method) \cite{binnig2014distributed} and \emph{Clock-SI} \cite{du2013clock}, which represents the best existing schedulers to enforce SI in distributed database systems. DSI allows local transactions to be handled by local nodes, while it requires centralized coordination for global / distributed transactions. Clock-SI does not need centralized coordination, but utilizes synchronized physical clocks to determine the time intervals of transactions. Time skew has negative impact on Clock-SI's performance. In the experiments, we used two versions of Clock-SI -- one uses completely synchronized clocks, representing the LAN setting (denoted by $Clock0$); the other represents the WAN setting (denoted by $Clock20$), to which we introduce a random time skew between -20ms and 20ms.
\end{enumerate}


To evaluate the performance, we used the benchmark of TPC-C and SmallBank. To store a database in our KV-store, we use a separate key-value hash structure to store each table. Each tuple is treated as a key-value pair -- the primary key of a tuple is treated as the key and the remaining part as the value. To support non-primary key queries, which are frequently invoked in TPC-C, we use secondary hash indexes. To conducts experiments on the aforementioned cluster with 30 virtual machines, we used one machine as the master and the others as slaves. The master does not participate in transaction processing. But it works as the central coordinator for SI and DSI, and the central point for clock synchronization for Clock-SI. Unless otherwise mentioned, each slave hosts 8 worker threads, which are fully devoted to transaction processing. For the tests on TPC-C, we installed 5 warehouses on each node. For the tests on SmallBank, we set the scale factor per node to 1 million customers. Transactions are divided into local transactions and distributed transactions. Each local transaction accesses only local data. Each distributed transaction accesses data from 2-3 randomly selected nodes.

\begin{figure}[t]
\centering
\includegraphics[width=.24\textwidth]{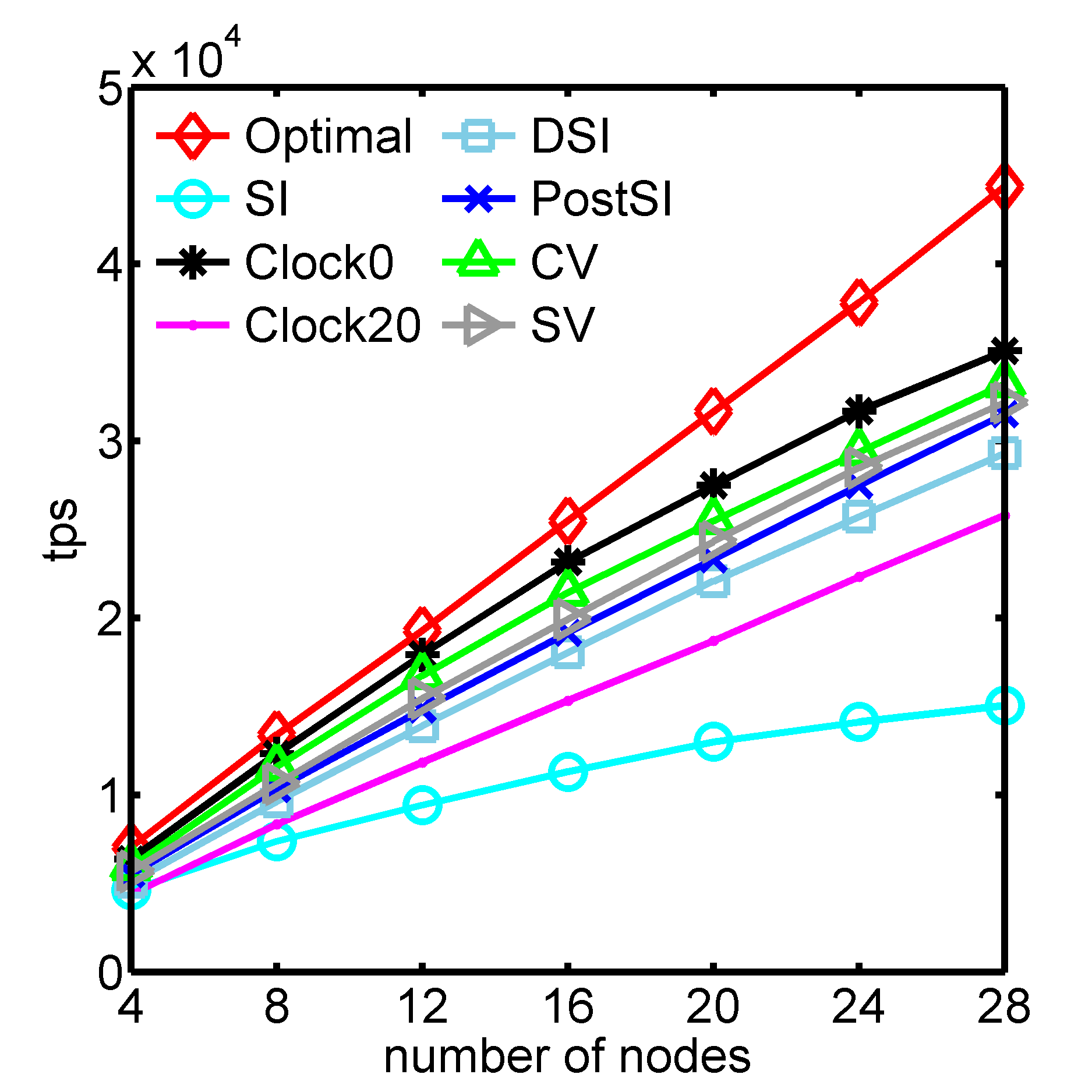}
\includegraphics[width=.24\textwidth]{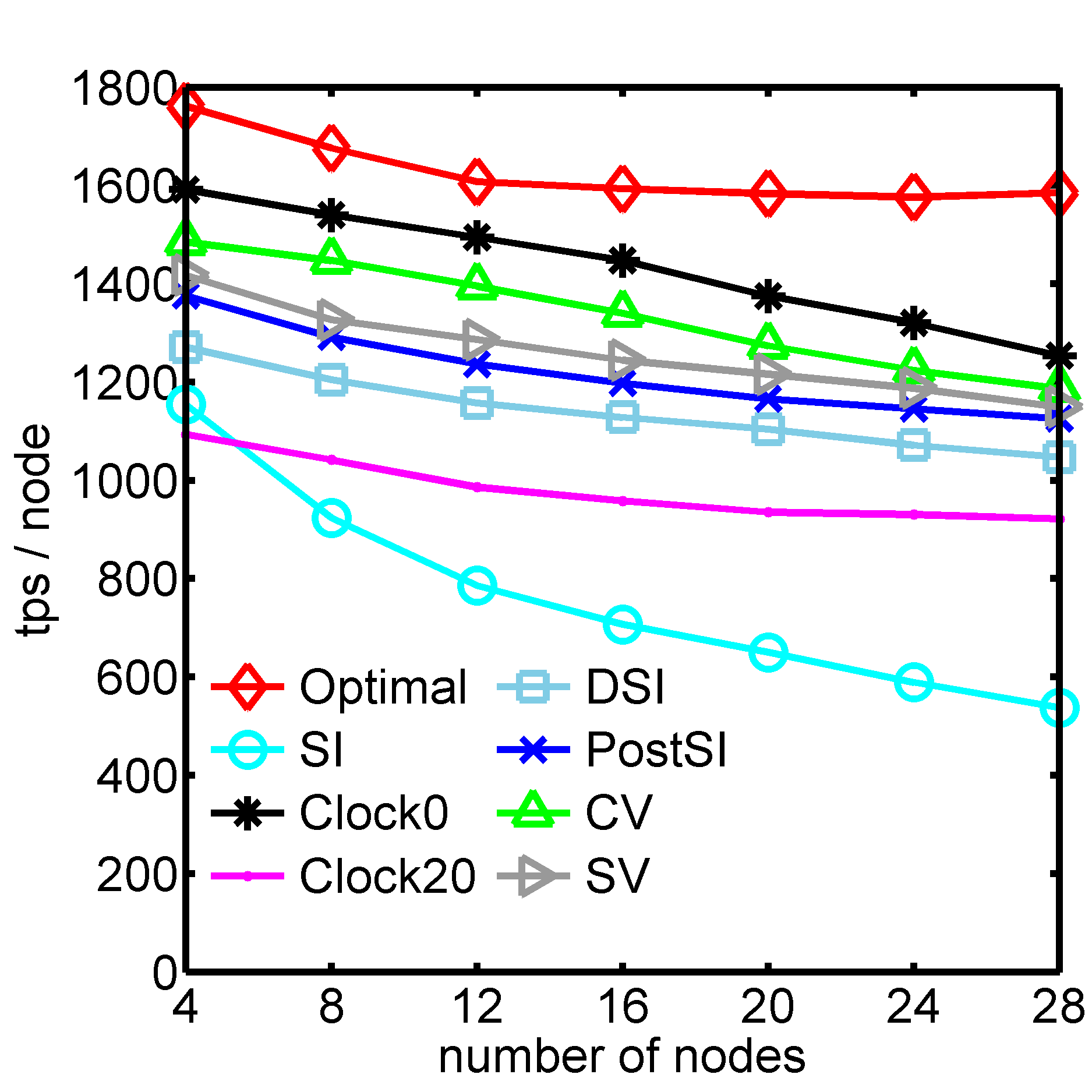}
\vspace{-2mm}
\caption{TPC-C Performance (20\% distributed txns)}
\label{fig:sn-20-tpcc}
\vspace{-3mm}
\end{figure}

\begin{figure}[t]
\centering
\includegraphics[width=.24\textwidth]{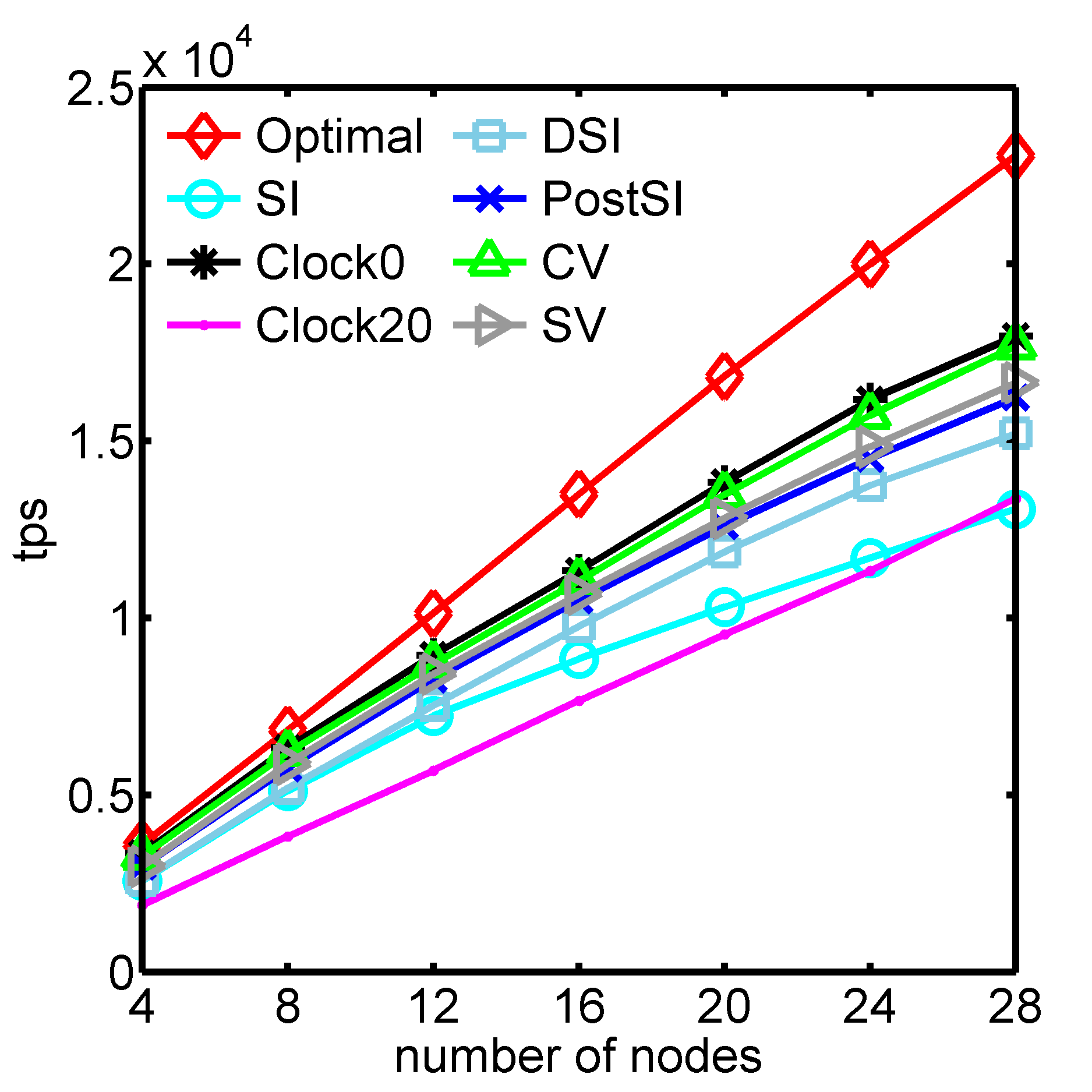}
\includegraphics[width=.24\textwidth]{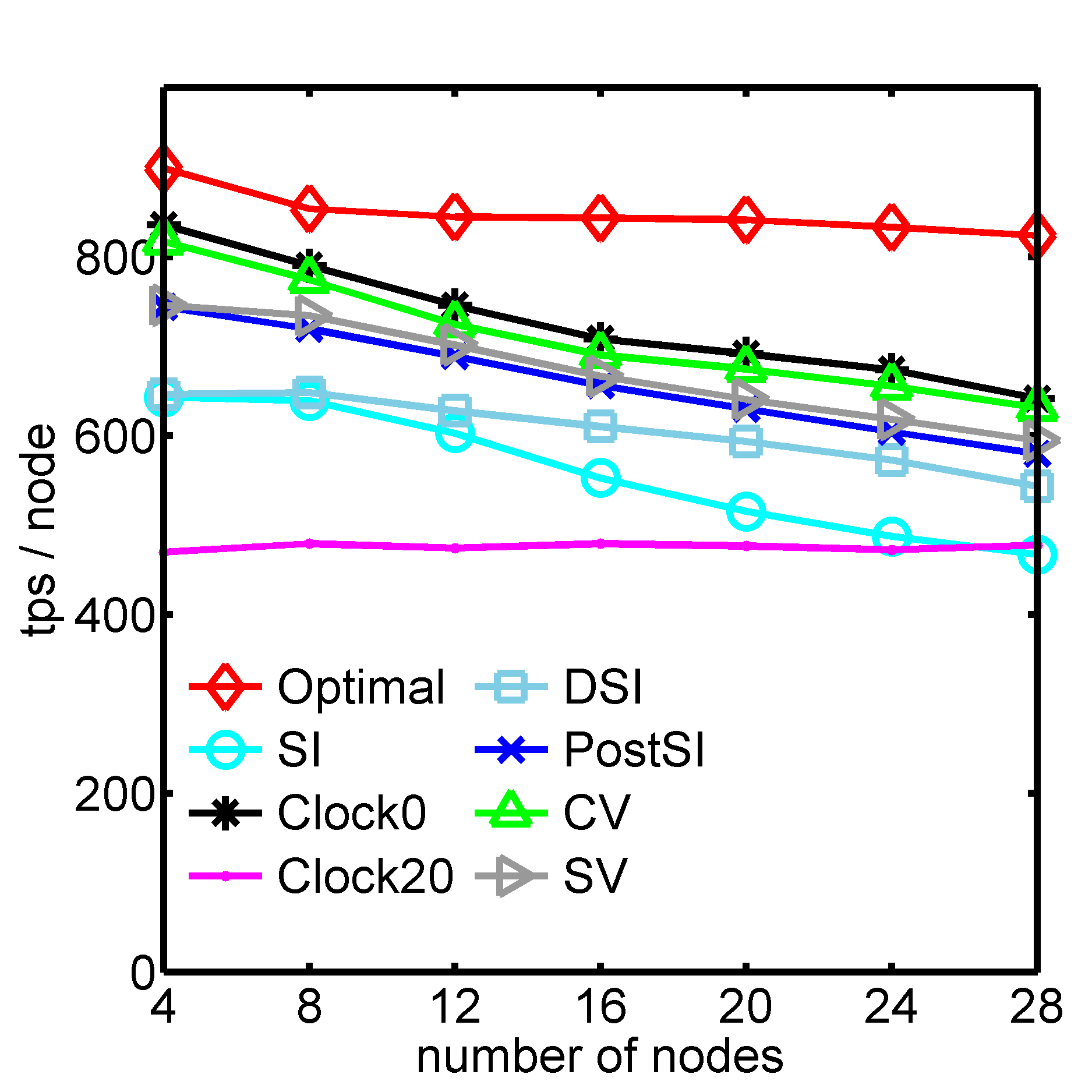}
\vspace{-2mm}
\caption{TPC-C Performance (50\% distributed txns)}
\label{fig:sn-20-small}
\vspace{-3mm}
\end{figure}

\begin{figure}[t]
\centering
\includegraphics[width=.24\textwidth]{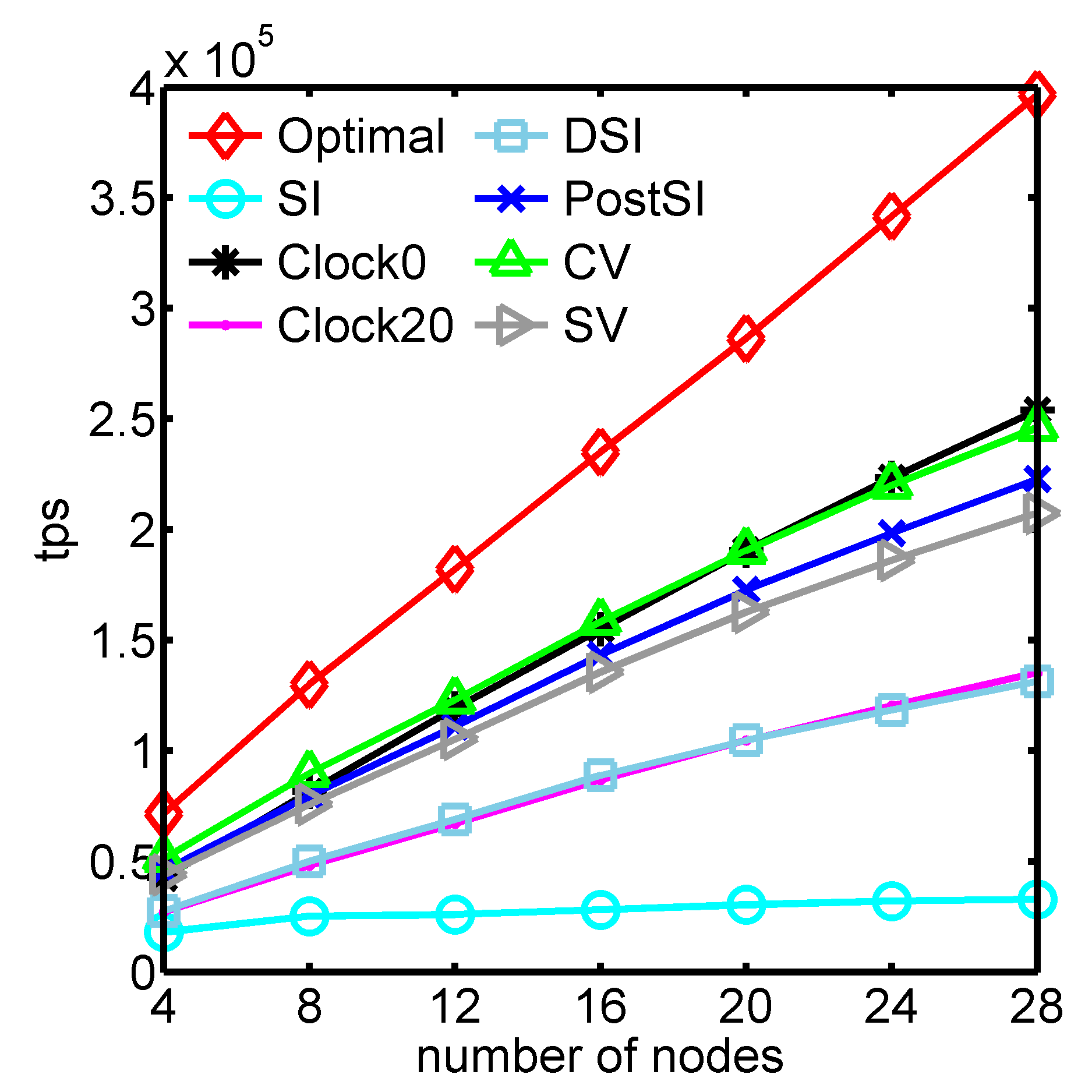}
\includegraphics[width=.24\textwidth]{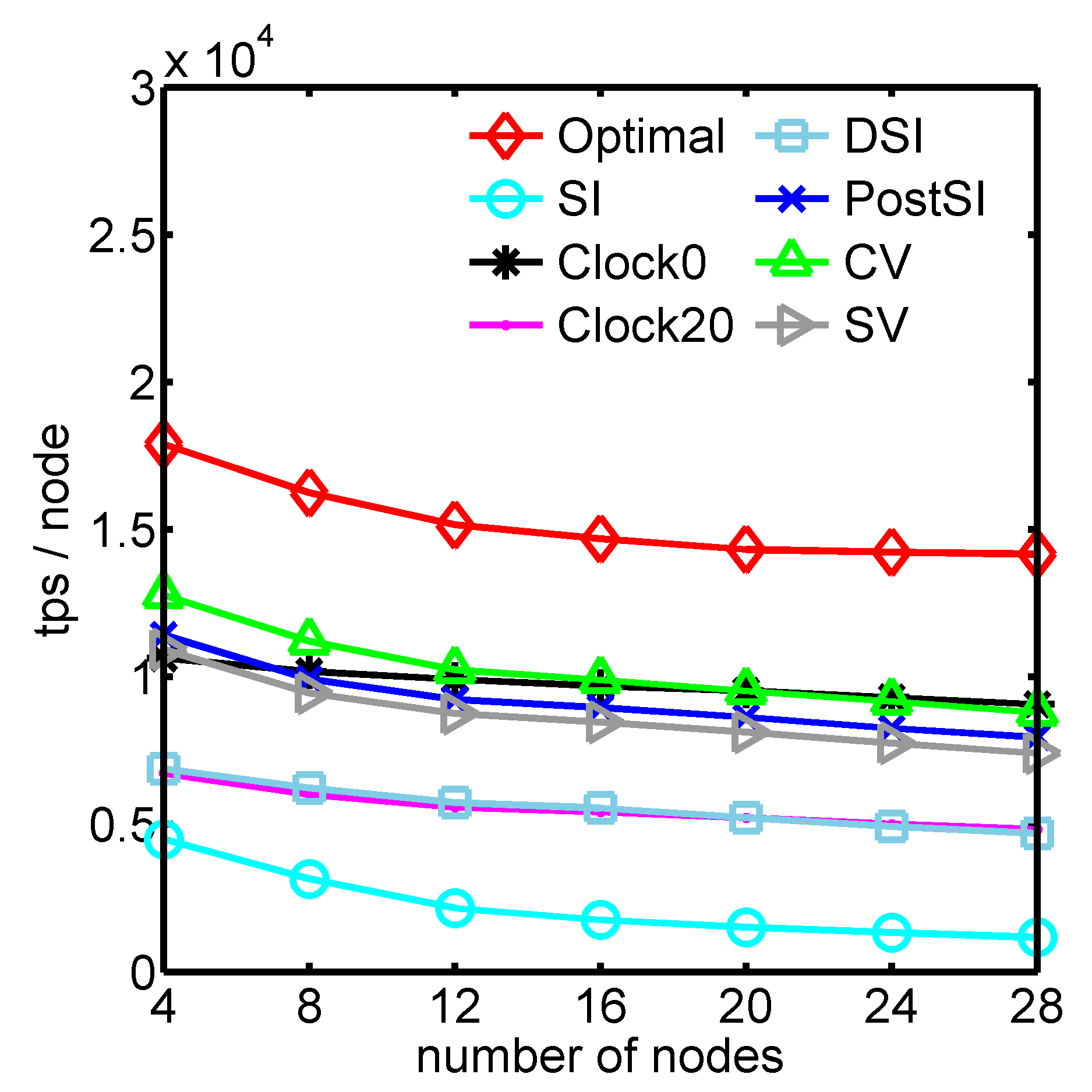}
\vspace{-2mm}
\caption{SmallBank Performance (20\% distributed txns)}
\label{fig:sn-50-tpcc}
\vspace{-3mm}
\end{figure}

\begin{figure}[t]
\centering
\includegraphics[width=.24\textwidth]{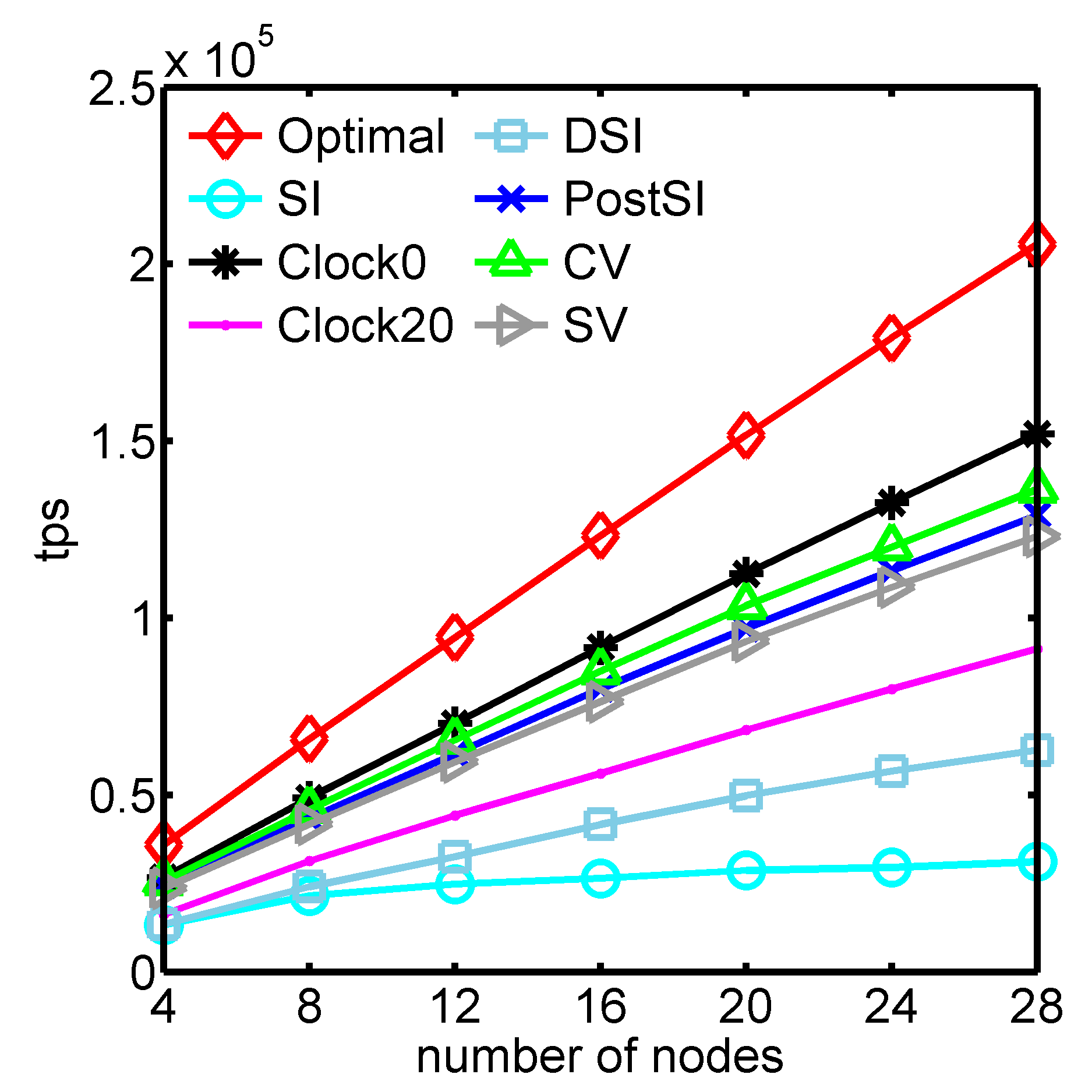}
\includegraphics[width=.24\textwidth]{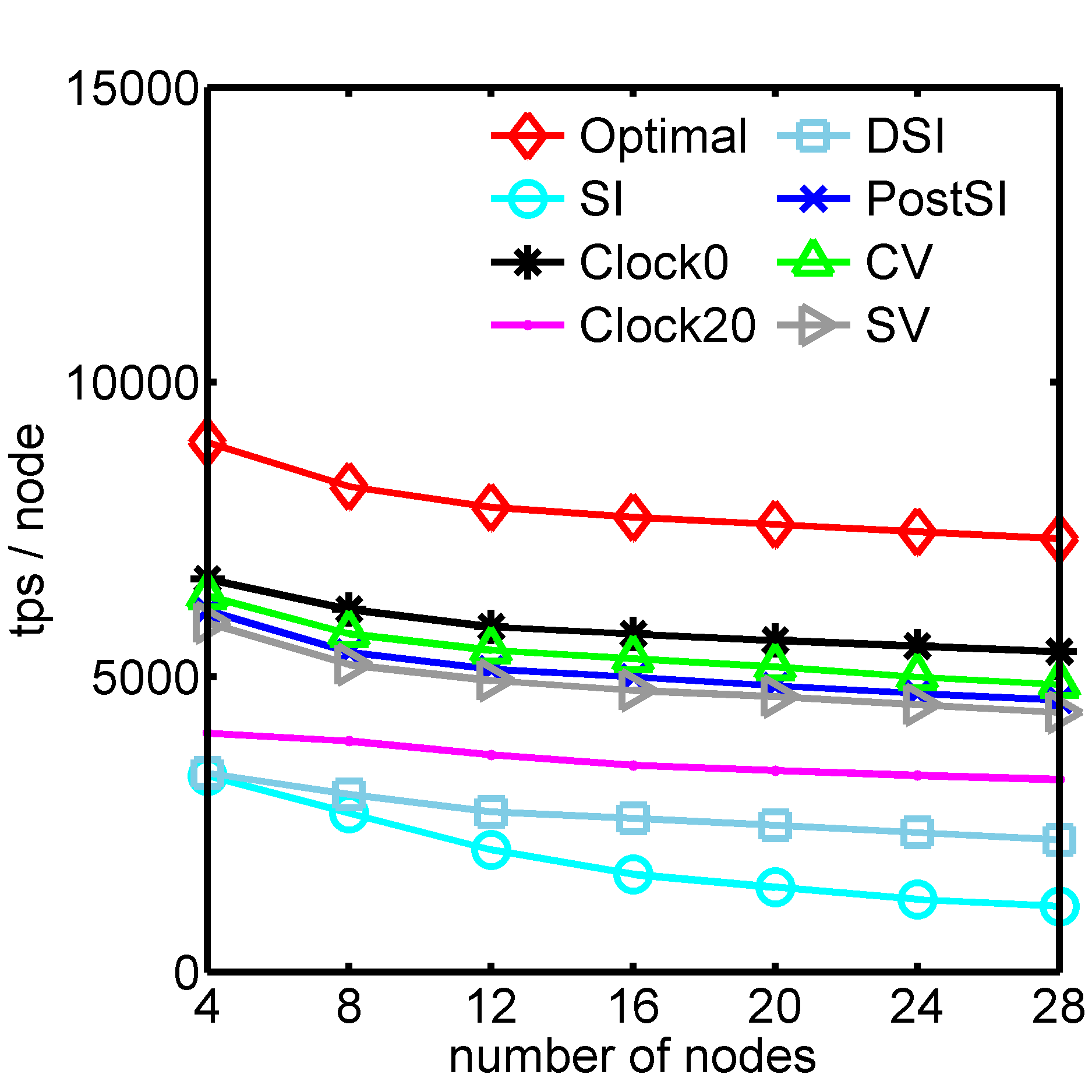}
\vspace{-2mm}
\caption{SmallBank Performance (50\% distributed txns)}
\label{fig:sn-50-small}
\vspace{-3mm}
\end{figure}



\subsection{Performance on Standard Benchmarks}

Our first set of experiments aimed to find out the general performance of the various schedulers on the benchmarks of TPC-C and Smallbank. In the experiments, we varied the number of slaves participating in the test, to study the scalability of the various schedulers. We also varied the proportion of distributed transactions (from 20\% to 50\%), to see how it influences the overall performance. Figures~\ref{fig:sn-20-tpcc}-\ref{fig:sn-50-small} show the performance of the various schedulers.

As we can see, the scalability of the conventional SI is the worst among all the schedulers. The growth of its performance clearly slows down when the number of the nodes reaches 16. At this point, the master is about to be saturated by the requests from the slaves, as each transaction requires two rounds of communication with the master. In contrast, the optimal scheduler performs the best, as it requires the least inter-node communication. CV, PostSI, SV, DSI and Clock-SI all outperform the conventional SI, while none of them can beat the optimal scheduler.

SmallBank contains simpler and shorter transactions than TPC-C. In our experiments, it also incurs less contention on data. As a result, the system's throughput on SmallBank is much higher than that on TPC-C. For schedulers such as DSI and conventional SI, a large proportion of their overheads is caused by the communication with the central coordinator, which occurs more frequently when transactions are shorter and run faster. Therefore, they perform worse in SmallBank than in TPC-C. In contrast, the schedulers of Clock-SI, CV, PostSI and SV are less affected by the length of transactions. 

The performance of Clock-SI can approaches that of the optimal scheduler, if the clocks of different nodes are completely synchronized. However, when there is time skew, its performance drops dramatically. Time skew hinders its performance in two ways -- first, the latency increases on the nodes that are ahead of the time, as they need to wait for the nodes that fall behind; second, the abort rate increases on the nodes that fall behind, because they often have to read older versions of the data. Especially when there is high contention, time skew can cause very high abort rate. As a result, $Clock20$ performs the worst in TPC-C.
Through an extra set of experiments, we found that Clock-SI's performance approaches that of PostSI if the time skew is limited to 5ms. In other words, $Clock5$ should perform as well as PostSI. However, controlling the time skew does not seem trivial.
True time devices, such as GPS clocks and atomic clocks, can be used to minimize the time skew and ensure the performance and stability of Clock-SI. This is out of the scope of this paper.

The performance of DSI is mainly influenced by distributed transactions. First, every distributed transaction needs to communicate with the central coordinator, which can be eventually saturated by an increasing number of requests. Second, DSI's abort rate on distributed transactions is usually high, as a mismatch between a local timestamp and the global timestamp can cause abort. This is confirmed by its authors in \cite{binnig2014distributed}. As shown in the results of SmallBank, when there are a large number of short distributed transactions, the scalability of DSI starts to hit a wall. In the experiments of TPC-C, as the transactions are significantly lengthier, the bottleneck of coordination is not yet visible. Therefore, DSI's performance is similar to that of PostSI on TPC-C.

\begin{figure}[t]
\centering
\includegraphics[width=.24\textwidth]{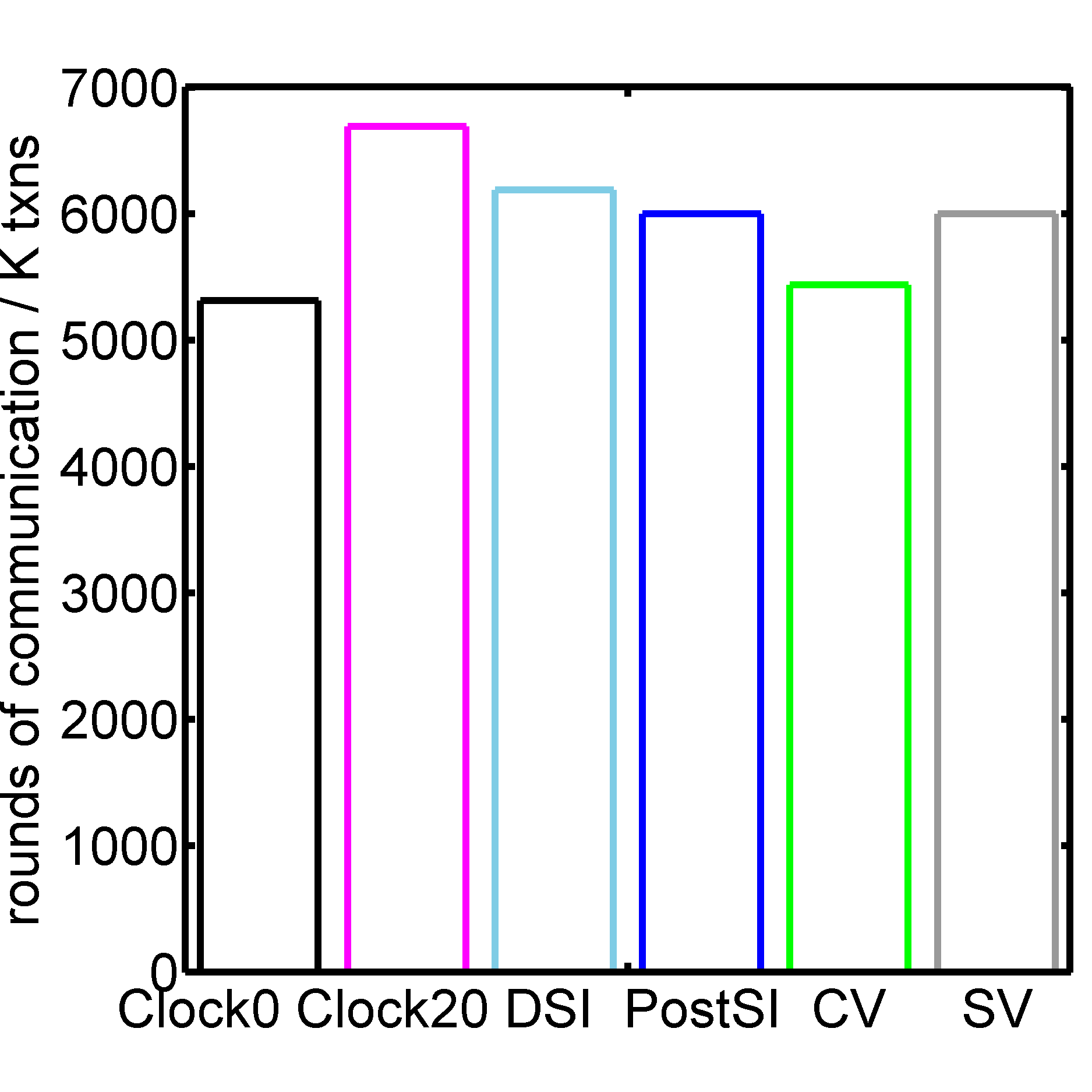}
\includegraphics[width=.24\textwidth]{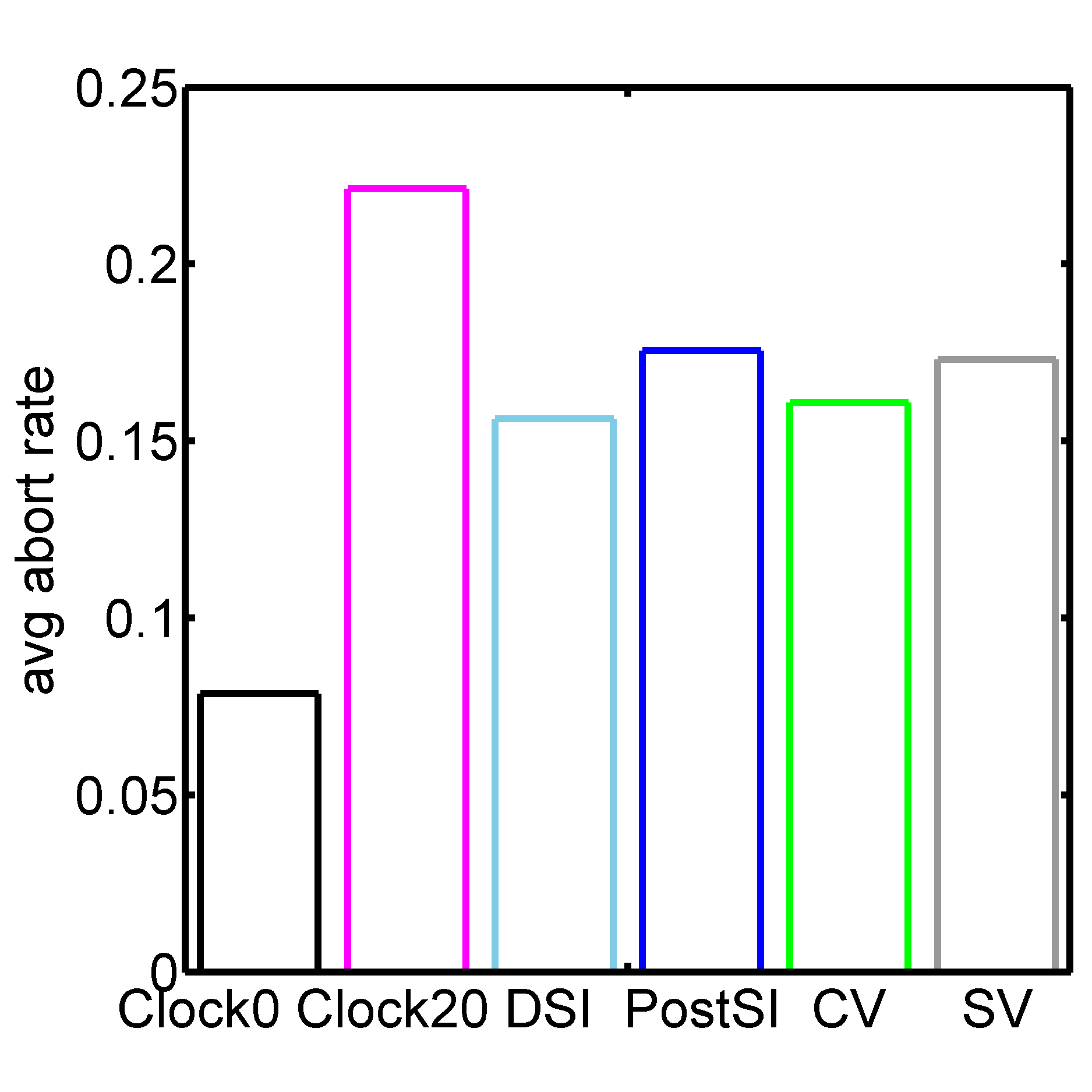}
\vspace{-2mm}
\caption{Communication Cost and Abort Rate (TPC-C, 8 nodes, 20\% distributed txns)}
\label{fig:commu-abort}
\vspace{-3mm}
\end{figure}

The ViCC schedulers, including CV, PostSI and SV, outperform DSI and the $Clock20$ in most of our experiments. The overheads of PostSI and SV are mainly caused by the communication for negotiating the orders of contending transactions. The overhead of CV is mainly caused by the transmission of anti-dependency information among the nodes. As the experiment results show, such extra communication is usually limited, as it only occurs when the contention between distributed transactions is high. All the three schedulers appear to scale well in TPC-C and SmallBank. CV performs slightly better than PostSI and SV, as it is weaker than PostSI and SV in isolation and allows for more concurrency. PostSI and SV exhibit similar performance. SV is in principle stricter than PostSI. This means that its abort rate can be higher than that of PostSI. This can be seen in the experiments of SmallBank, in which SV performs slightly worse than PostSI because of the risen abort rates. However, in the experiments of TPC-C, this difference is invisible, as it has been proven that all SI schedules are serializable in TPC-C \cite{fekete2005making,Cahill2008}. In contrast, SV can sometimes perform better than PostSI in TPC-C, as SV only needs to maintain a single order $o$, which costs less than maintaining an time interval $(s,c)$.

For transaction management on an MPP platform, the frequency of cross-node communication seems to be one of the dominant factors for performance. The abort rates can also provide an insight about the cause of bad performance. Figure~\ref{fig:commu-abort} shows the communication cost and the abort rates of the various schedulers in a TPC-C test. It is consistent the performance results in Figure~\ref{fig:sn-20-tpcc}.

\subsection{Characteristics of the Schedulers}

We designed several additional sets of experiments to study the characteristics of the various schedulers.
We used Smallbank as the benchmark. We varied the degree of contention, the lengths of transactions and the fraction of distributed transactions in Smallbank, and observed how these factors influence the performance of the schedulers. All the experiments were conducted on 20 nodes.

\begin{figure}[t]
\centering
\includegraphics[width=.24\textwidth]{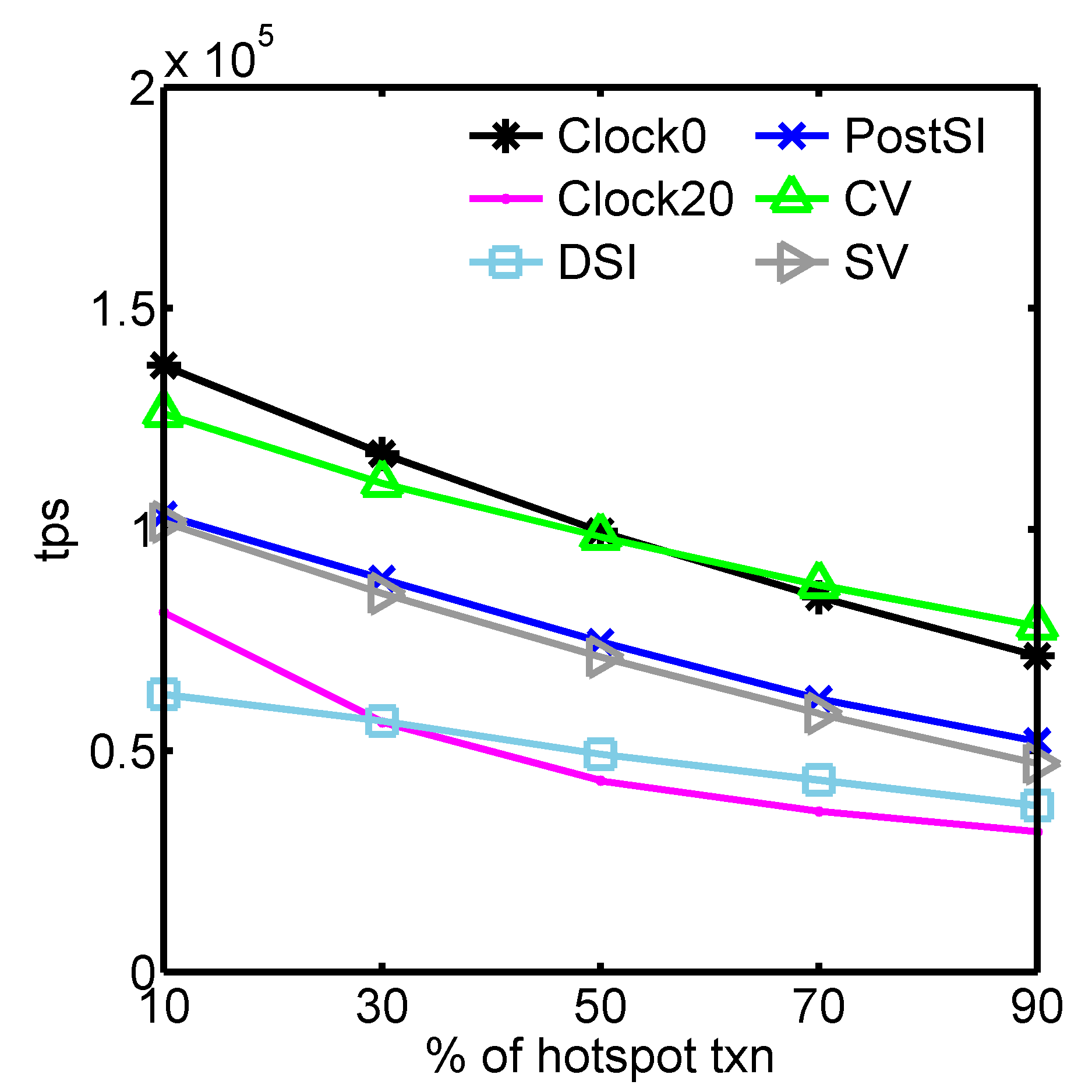}
\includegraphics[width=.24\textwidth]{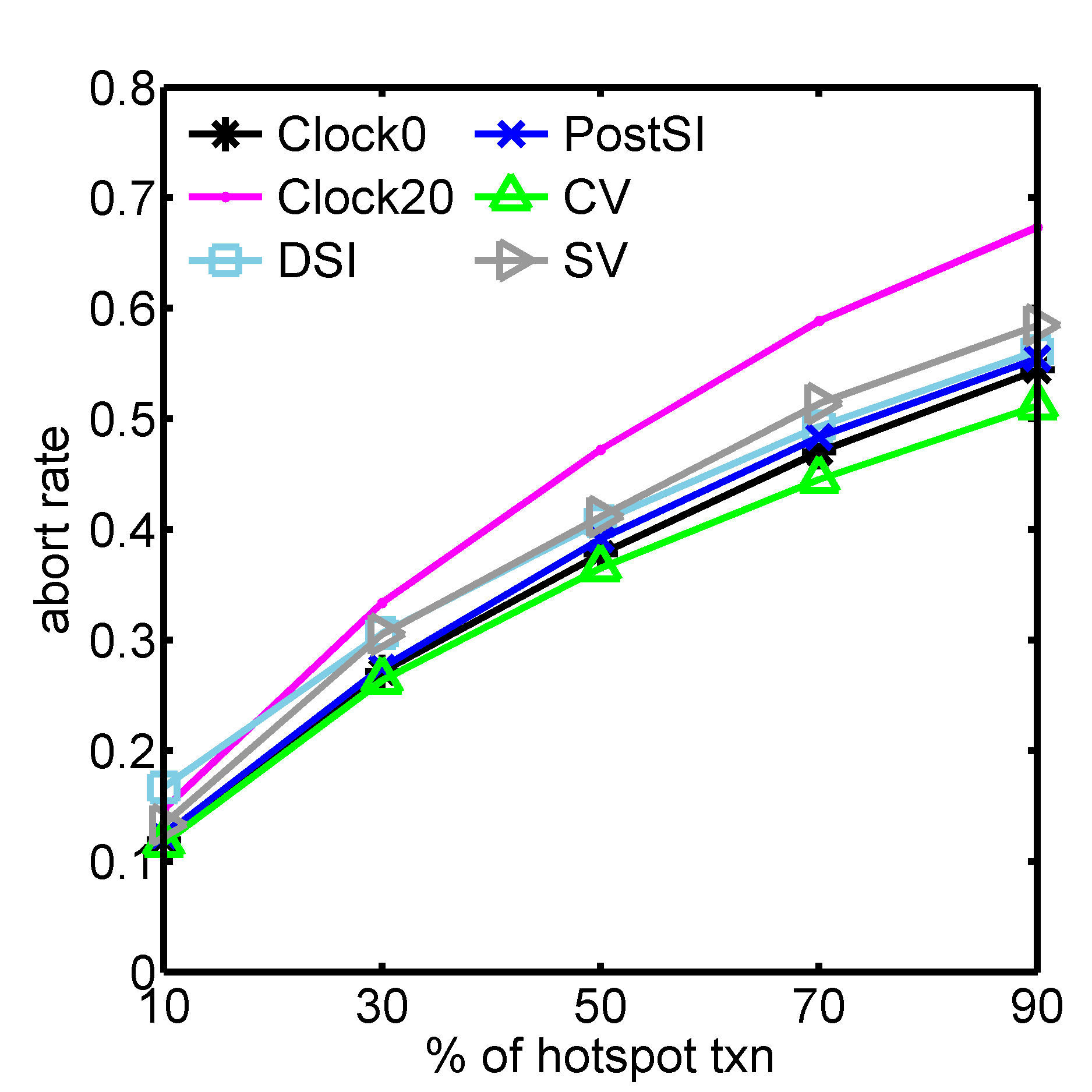}
\vspace{-2mm}
\caption{Varying Degree of Contention (30\% distributed txns)}
\label{fig:hotspots}
\vspace{-2mm}
\end{figure}

In the first set of experiments, we varied the degree of contention by varying the proportion of transactions that access hotspot data. (On each node, 20 out of the 1 million data items are classified as the hotspot.) Figure~\ref{fig:hotspots} shows how the various SI schedulers behave. As expected, when contention increases, the throughput of all the schedulers declines. This is mainly due to the rising abort rates caused by contention, which is shown on the right of Figure~\ref{fig:hotspots}. To CV, PostSI and SV, higher contention also incur higher communication cost, as more negotiation has to be carried out between transactions. As a result, the performance decline of the three schedulers appears slightly sharper than that of DSI. In contrast to the others, Clock-SI cannot guarantee non-blocking read -- when a transaction enters its commit phase, it will block the reads on the data it has updated. This increases Closk-SI's cost in dealing with contention. Therefore, the performance of Clock-SI declines even faster.

\begin{figure}[t]
\centering
\includegraphics[width=.24\textwidth]{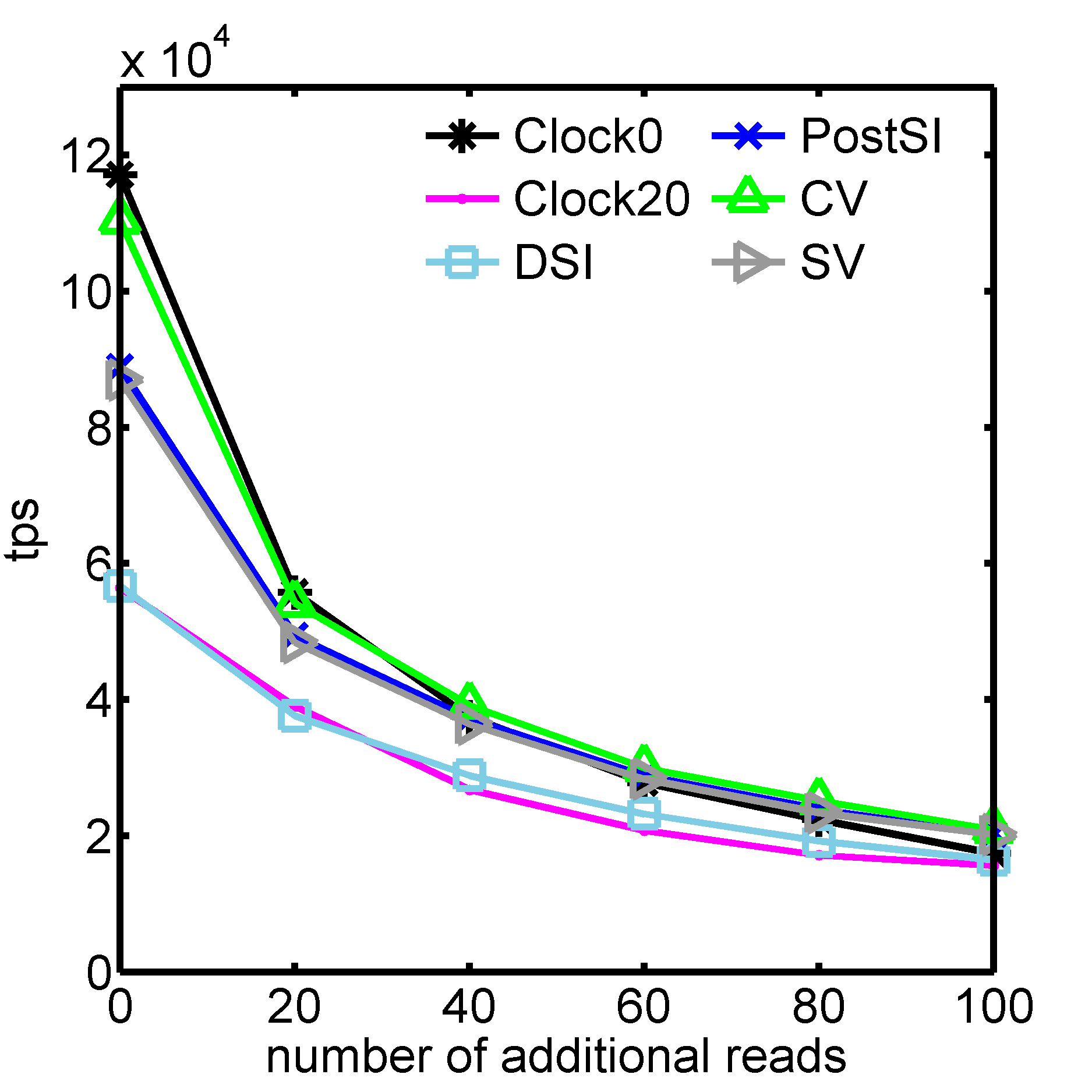}
\includegraphics[width=.24\textwidth]{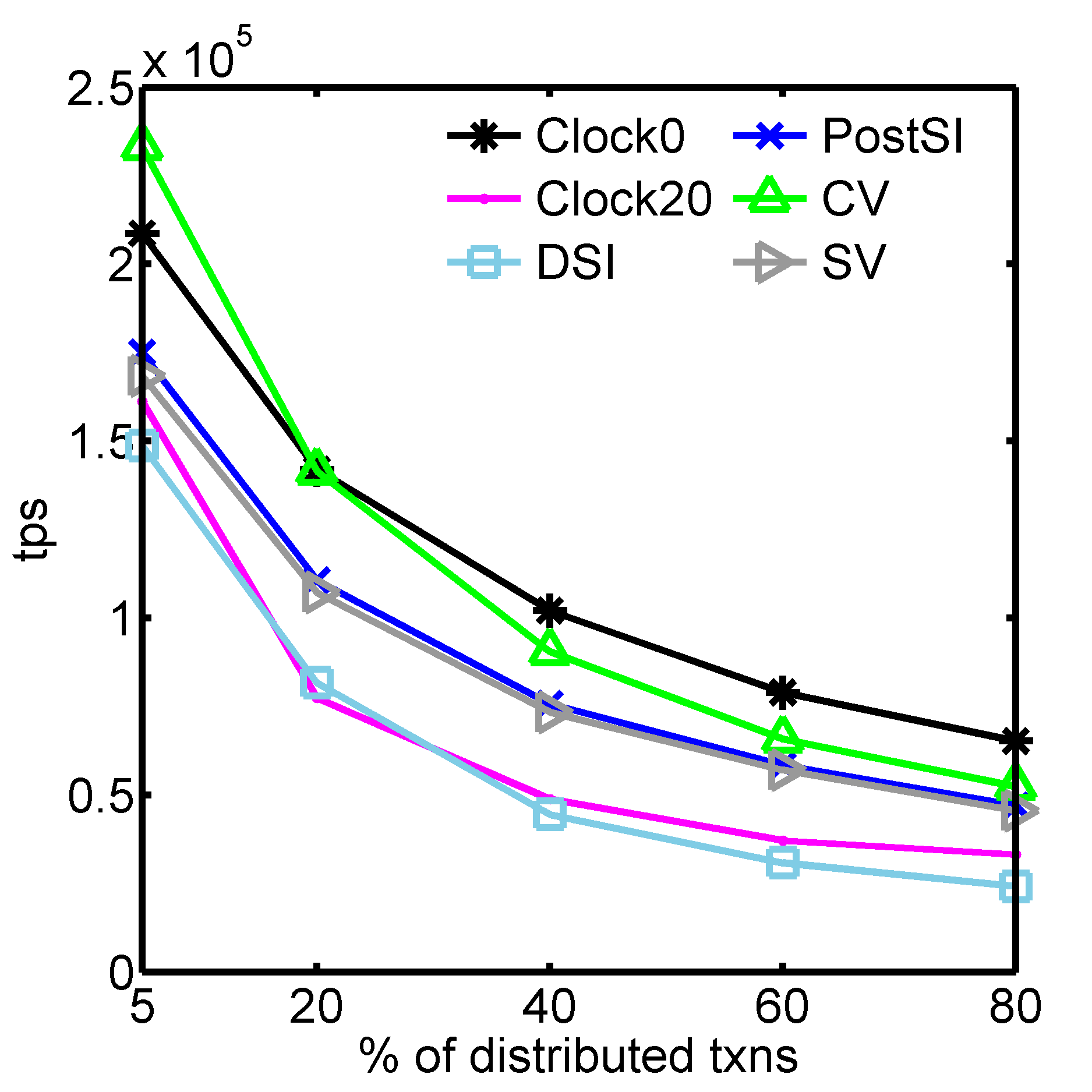}
(a)\hspace{0.22\textwidth}(b)
\caption{(a) Varying Length of Transaction (30\% distributed txns); (b) Varying Proportion of Distributed Transactions}
\label{fig:txnlength}
\vspace{-3mm}
\end{figure}

In the second set of experiments, we gradually increased the length of each transaction by adding random read operations to it. Figure~\ref{fig:txnlength}(a) shows that the performance gap between the schedulers drops as the transaction length increases. To most of the schedulers, the scheduling cost per transaction remains almost the same, regardless of the transaction length. When the transactions are longer, less transactions will be executed. As a result, the scheduling cost drops. This explains why the performance gap in TPC-C is smaller than that in Smallbank (Figures~\ref{fig:sn-20-tpcc}-\ref{fig:sn-50-small}).

In the third set of experiments, we varied the fraction of distributed transactions from 5\% to 80\%. The performance of the various schedulers are shown in Figure~\ref{fig:txnlength}(b). When there are more distributed transactions, more cross-node communication will occur. Thus, the performance of all the schedulers drops. CV's performance seems to drop slightly faster than the others. This is because CV needs to lookup the anti-dependency table when assessing the visibility of data, which incurs extra cross-node communication. As mentioned in Section~4.2, if we let CV to maintain a CID for each data object, this communication cost can actually be saved.

\subsection{Comparison Against TicToc}

The main difference between ViCC and TicToc lies in how they identify and deal with anti-dependencies (or read-write dependencies). ViCC employs an anti-dependency table; transactions performing the writes are responsible for recording anti-dependencies in the table. In contrast, TicToc adopts the OCC approach -- it checks anti-dependencies in the validation phase by examining the read set to see if it has been updated. To understand how the design choices influence the performance, we conducted an additional set of experiments.

\begin{figure}[t]
\centering
\includegraphics[width=.24\textwidth]{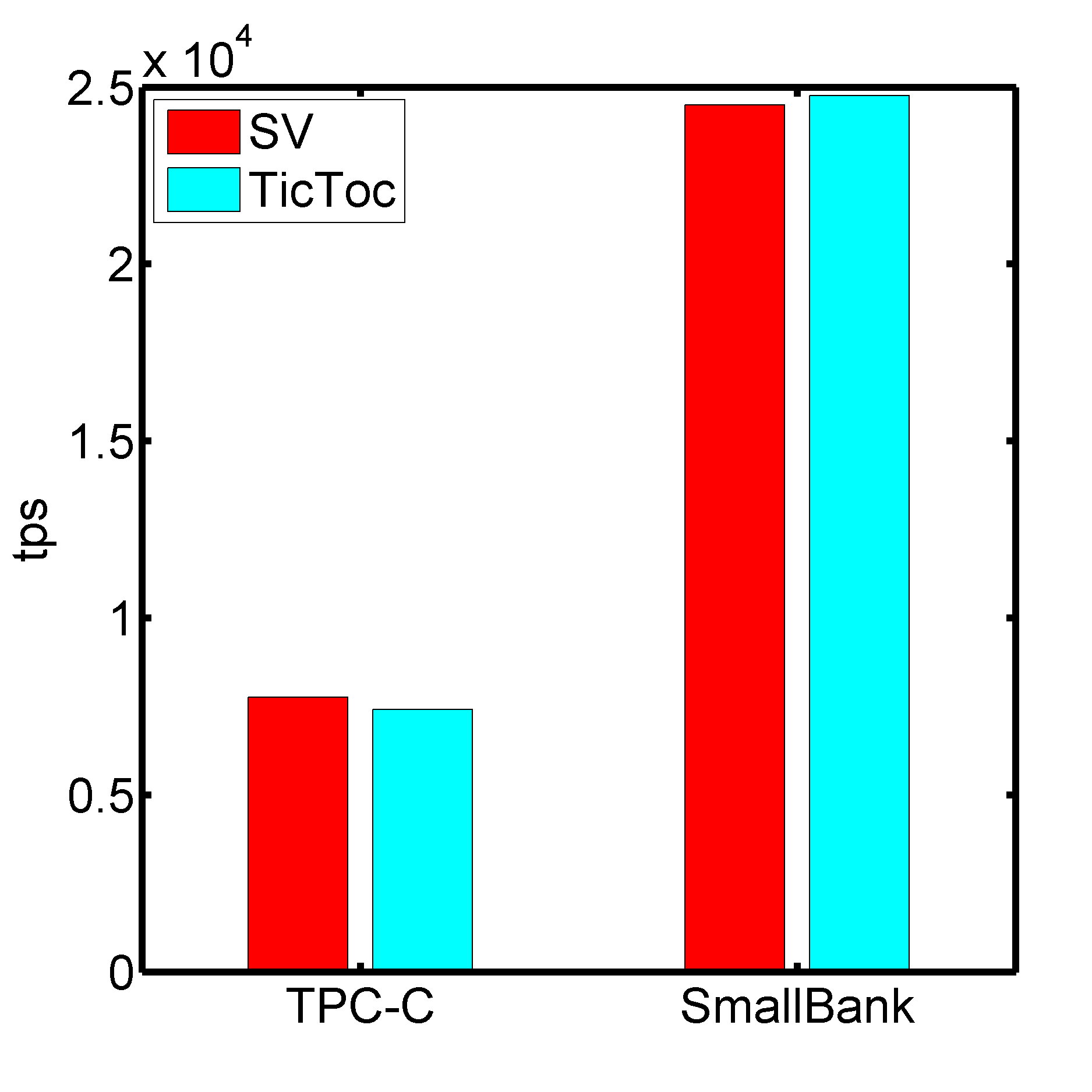}
\includegraphics[width=.24\textwidth]{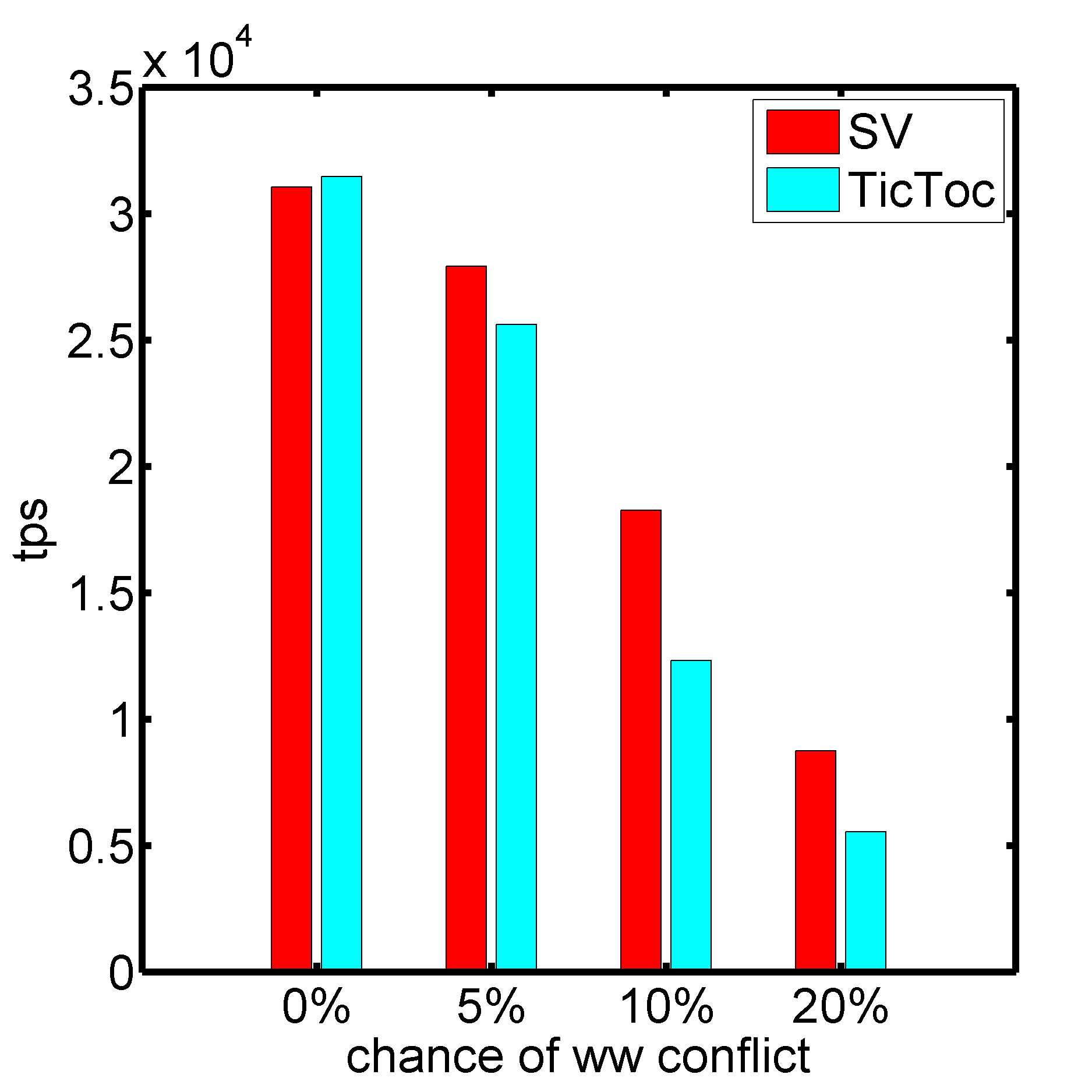}
(a) benchmarks ~~~~~~~~~~~~~~~~ (b) write contention
\caption{Performance Against TicToc }
\label{fig:tictoc-bench}
\vspace{-2mm}
\end{figure}


We implemented TicToc within the same framework of ViCC. TicToc requires each transaction to maintain a read set and a write set. We distribute the read and write sets to the data nodes, so that the insertion and validation of the read set can be performed locally on each node. 
Similar to PostSI and SV, TicToc performs three rounds of communication when finishing a transaction. In the first round, the host of the transaction informs the data nodes to lock the data in the write set and compute a global commit timestamp. In the second round, the host broadcasts the commit timestamp to the data nodes and starts the validation phase. In the final round, according to the outcome of the validation, the host informs the data nodes either to abort or the commit.

As TicToc is designed to enforce serializability, we mainly compared it against our SV scheduler. Our experiments were conducted on 8 nodes. Figure~\ref{fig:tictoc-bench}(a) shows the performance of SV and TicToc on TPC-C and Smallbank. TicToc demonstrates similar performance as SV on the benchmarks. While TicToc has to perform an additional validation step at the end, it does not need to update the visitor lists or the anti-dependency table as SV does. Both approaches seem to incur the same amount of communication cost.

To figure out their difference, we were compelled to design a special test. In the test, we created a single table consisting of 1 million tuples. We let each transaction randomly read 100 tuples and update 2 tuples. We varied the chance of write-write conflict by forcing a fraction of transactions to update a single hotspot tuple. Figure~\ref{fig:tictoc-bench}(b) show the performance of SV and TicToc. As we can see, when there is little contention, SV and TicToc perform similarly. However, when the chance of write-write conflict increases, the performance of TicToc drops faster than SV. It is because TicToc holds locks longer than SV does -- the validation step has be to performed when the locks are held. This also explains why TicToc's performance on TPC-C was slightly worse than SV's (Figure~\ref{fig:tictoc-bench}(a)), as the \emph{next order id} of TPC-C is a hotspot of update. Thus, we can conclude that ViCC can be superior to TicToc when there is a hotspot of update.

\section{Conclusion}

In this paper, we introduced Visibility based Concurrency Control (ViCC), a series of concurrency control mechanisms that do not require centralized coordination.
Instead of relying on a central clock, ViCC allows transactions to determine their temporal orders autonomously, by monitoring and negotiating on their visibility relationships. We introduced three isolation levels for ViCC, named Consistent Visibility (CV), Posterior Snapshot Isolation (PostSI) and Serializable Visibility (SV). CV requires that the visibility between each pair of transactions to be atomic. PostSI and SV are identical to traditional Snapshot Isolation and Serializablity. They impose additional constraints on the orders of visibility. As the PostSI and SV schedulers build upon the CV scheduler, the three schedulers can be built into a single coherent mechanism. This makes them practical to implement.
Through the concept of visibility, we proved the soundness of the schedulers. Extensive experimental evaluation demonstrated their suitability for shared-nothing architectures.
Our future research will explore the ways to integrate a replication scheme into ViCC, so that our schedulers can be utilized by real world distributed databases.
\\




\bibliographystyle{IEEEtran}
\bibliography{dsi-icde-arxiv}

\begin{thebibliography}{10}
\providecommand{\url}[1]{#1}
\csname url@samestyle\endcsname
\providecommand{\newblock}{\relax}
\providecommand{\bibinfo}[2]{#2}
\providecommand{\BIBentrySTDinterwordspacing}{\spaceskip=0pt\relax}
\providecommand{\BIBentryALTinterwordstretchfactor}{4}
\providecommand{\BIBentryALTinterwordspacing}{\spaceskip=\fontdimen2\font plus
\BIBentryALTinterwordstretchfactor\fontdimen3\font minus
  \fontdimen4\font\relax}
\providecommand{\BIBforeignlanguage}[2]{{%
\expandafter\ifx\csname l@#1\endcsname\relax
\typeout{** WARNING: IEEEtran.bst: No hyphenation pattern has been}%
\typeout{** loaded for the language `#1'. Using the pattern for}%
\typeout{** the default language instead.}%
\else
\language=\csname l@#1\endcsname
\fi
#2}}
\providecommand{\BIBdecl}{\relax}
\BIBdecl

\bibitem{oracleSI}
Oracle, ``Chapter 9: Data concurrency and consistency ¡ª overview of {Oracle}
  database transaction isolation levels,'' in \emph{Oracle Database Concepts
  11g Release 2 (11.2)}, 2015, ch.~9.

\bibitem{SQLServerSI}
Microsoft, ``Snapshot isolation in {SQL} {Server},'' in \emph{.NET Framework
  4.6 and 4.5}, 2015.

\bibitem{ports2012serializable}
D.~R. Ports and K.~Grittner, ``Serializable snapshot isolation in
  {PostgreSQL},'' \emph{PVLDB}, vol.~5, no.~12, pp. 1850--1861, 2012.

\bibitem{berenson1995critique}
H.~Berenson, P.~Bernstein, J.~Gray, J.~Melton, E.~O'Neil, and P.~O'Neil, ``A
  critique of ansi sql isolation levels,'' \emph{ACM SIGMOD Record}, vol.~24,
  no.~2, pp. 1--10, 1995.

\bibitem{corbett2013spanner}
J.~C. Corbett, J.~Dean, M.~Epstein, A.~Fikes, C.~Frost, J.~J. Furman,
  S.~Ghemawat, A.~Gubarev, C.~Heiser, P.~Hochschild \emph{et~al.}, ``Spanner:
  Google¡¯s globally distributed database,'' \emph{ACM TOCS}, vol.~31, no.~3,
  p.~8, 2013.

\bibitem{schulz1999end}
M.~Schulz, ``The end of the road for silicon?'' \emph{Nature}, vol. 399, no.
  6738, pp. 729--730, 1999.

\bibitem{kish2002end}
L.~B. Kish, ``End of {Moore}'s law: thermal (noise) death of integration in
  micro and nano electronics,'' \emph{Physics Letters A}, vol. 305, no.~3, pp.
  144--149, 2002.

\bibitem{pavlo15}
A.~Pavlo, ``Emerging hardware trends in large-scale transaction processing,''
  \emph{IEEE Internet Computing}, vol.~19, no.~3, pp. 68--71, May/June 2015.

\bibitem{yu2014}
X.~Yu, G.~Bezerra, A.~Pavlo, S.~Devadas, and M.~Stonebraker, ``Staring into the
  abyss: An evaluation of concurrency control with one thousand cores,''
  \emph{PVLDB}, vol.~8, pp. 209--220, November 2014.

\bibitem{cattell2011scalable}
R.~Cattell, ``Scalable sql and nosql data stores,'' \emph{ACM SIGMOD Record},
  vol.~39, no.~4, pp. 12--27, 2011.

\bibitem{stonebraker2012newsql}
M.~Stonebraker, ``Newsql: An alternative to nosql and old sql for new oltp
  apps,'' \emph{Communications of the ACM. Retrieved}, pp. 07--06, 2012.

\bibitem{zhou2017}
X.~Zhou, X.~Zhou, Z.~Yu, and K.~L. Tan, ``Posterior snapshot isolation,'' in
  \emph{ICDE}, 2017, pp. 797--808.

\bibitem{Bernstein:1981:CCD}
P.~A. Bernstein and N.~Goodman, ``Concurrency control in distributed database
  systems,'' \emph{ACM Comput. Surv.}, vol.~13, no.~2, pp. 185--221, Jun. 1981.

\bibitem{weikum2001transactional}
G.~Weikum and G.~Vossen, \emph{Transactional information systems: theory,
  algorithms, and the practice of concurrency control and recovery}.\hskip 1em
  plus 0.5em minus 0.4em\relax Elsevier, 2001.

\bibitem{Harding:2017:EDC}
R.~Harding, D.~Van~Aken, A.~Pavlo, and M.~Stonebraker, ``An evaluation of
  distributed concurrency control,'' \emph{PVLDB}, vol.~10, no.~5, pp.
  553--564, 2017.

\bibitem{DBLP:conf/usenix/CowlingL12}
J.~A. Cowling and B.~Liskov, ``Granola: Low-overhead distributed transaction
  coordination,'' in \emph{USENIX ATC}, 2012, pp. 223--235.

\bibitem{Thomasian:1998:DOC}
A.~Thomasian, ``Distributed optimistic concurrency control methods for
  high-performance transaction processing,'' \emph{IEEE TKDE}, vol.~10, no.~1,
  pp. 173--189, Jan. 1998.

\bibitem{Mu:2014:EMC}
S.~Mu, Y.~Cui, Y.~Zhang, W.~Lloyd, and J.~Li, ``Extracting more concurrency
  from distributed transactions,'' in \emph{OSDI}, 2014, pp. 479--494.

\bibitem{Bernstein:1986:CCR:17299}
P.~A. Bernstein, V.~Hadzilacos, and N.~Goodman, \emph{Concurrency Control and
  Recovery in Database Systems}.\hskip 1em plus 0.5em minus 0.4em\relax Boston,
  MA, USA: Addison-Wesley Longman Publishing Co., Inc., 1986.

\bibitem{schenkel2000federated}
R.~Schenkel, G.~Weikum, N.~Wei{\ss}enberg, and X.~Wu, ``Federated transaction
  management with snapshot isolation,'' in \emph{Transactions and Database
  Dynamics}, 2000, pp. 1--25.

\bibitem{binnig2014distributed}
C.~Binnig, S.~Hildenbrand, F.~F{\"a}rber, D.~Kossmann, J.~Lee, and N.~May,
  ``Distributed snapshot isolation: global transactions pay globally, local
  transactions pay locally,'' \emph{The VLDB Journal}, vol.~23, no.~6, pp.
  987--1011, 2014.

\bibitem{sovran2011transactional}
Y.~Sovran, R.~Power, M.~K. Aguilera, and J.~Li, ``Transactional storage for
  geo-replicated systems,'' in \emph{SOSP}, 2011, pp. 385--400.

\bibitem{zhang2010supporting}
C.~Zhang and H.~De~Sterck, ``Supporting multi-row distributed transactions with
  global snapshot isolation using bare-bones hbase,'' in \emph{International
  Conference on Grid Computing}, 2010, pp. 177--184.

\bibitem{zhang2011hbasesi}
C.~Zhang and H.~D. Sterck, ``Hbasesi: Multi-row distributed transactions with
  global strong snapshot isolation on clouds,'' \emph{Scalable Computing:
  Practice and Experience}, vol.~12, no.~2, 2011.

\bibitem{lee2013sap}
J.~Lee, Y.~S. Kwon, F.~Farber, M.~Muehle, C.~Lee, C.~Bensberg, J.~Y. Lee, A.~H.
  Lee, and W.~Lehner, ``Sap hana distributed in-memory database system:
  Transaction, session, and metadata management,'' in \emph{ICDE}, 2013, pp.
  1165--1173.

\bibitem{du2013clock}
J.~Du, S.~Elnikety, and W.~Zwaenepoel, ``Clock-si: Snapshot isolation for
  partitioned data stores using loosely synchronized clocks,'' in \emph{SRDS},
  2013, pp. 173--184.

\bibitem{ardekani2013non}
M.~S. Ardekani, P.~Sutra, and M.~Shapiro, ``Non-monotonic snapshot isolation:
  scalable and strong consistency for geo-replicated transactional systems,''
  in \emph{SRDS}, 2013, pp. 163--172.

\bibitem{tripathi2015scalable}
A.~Tripathi, G.~Rajappan, and V.~Padhye, ``Scalable transactions in partially
  replicated data systems with causal snapshot isolation,'' \emph{Technical
  Report}, 2015.

\bibitem{adya1995efficient}
A.~Adya, R.~Gruber, B.~Liskov, and U.~Maheshwari, ``Efficient optimistic
  concurrency control using loosely synchronized clocks,'' \emph{ACM SIGMOD
  Record}, vol.~24, no.~2, pp. 23--34, 1995.

\bibitem{lomet2012multi}
D.~Lomet, A.~Fekete, R.~Wang, and P.~Ward, ``Multi-version concurrency via
  timestamp range conflict management,'' in \emph{ICDE}, 2012, pp. 714--725.

\bibitem{yu2016tictoc}
X.~Yu, A.~Pavlo, D.~Sanchez, and S.~Devadas, ``Tictoc: Time traveling
  optimistic concurrency control,'' in \emph{SIGMOD}, vol.~8, 2016, pp.
  209--220.

\bibitem{Faleiro:2015:RSM}
J.~M. Faleiro and D.~J. Abadi, ``Rethinking serializable multiversion
  concurrency control,'' \emph{PVLDB}, vol.~8, no.~11, pp. 1190--1201, Jul.
  2015.

\bibitem{Neumann:2015:FSM}
T.~Neumann, T.~M\"{u}hlbauer, and A.~Kemper, ``Fast serializable multi-version
  concurrency control for main-memory database systems,'' in \emph{SIGMOD},
  2015, pp. 677--689.

\bibitem{Dashti:2017:TRM}
M.~Dashti, S.~Basil~John, A.~Shaikhha, and C.~Koch, ``Transaction repair for
  multi-version concurrency control,'' in \emph{SIGMOD}, 2017, pp. 235--250.

\bibitem{elnikety2005database}
S.~Elnikety, F.~Pedone, and W.~Zwaenepoel, ``Database replication using
  generalized snapshot isolation,'' in \emph{SRDS}, 2005, pp. 73--84.

\bibitem{daudjee2006lazy}
K.~Daudjee and K.~Salem, ``Lazy database replication with snapshot isolation,''
  in \emph{VLDB}, 2006, pp. 715--726.

\bibitem{jung2011serializable}
H.~Jung, H.~Han, A.~Fekete, and U.~R{\"o}hm, ``Serializable snapshot isolation
  for replicated databases in high-update scenarios,'' \emph{PVLDB}, vol.~4,
  no.~11, pp. 783--794, 2011.

\bibitem{lin2009snapshot}
Y.~Lin, B.~Kemme, R.~Jim{\'e}nez-Peris, M.~Pati{\~n}o-Mart{\'\i}nez, and J.~E.
  Armend{\'a}riz-I{\~n}igo, ``Snapshot isolation and integrity constraints in
  replicated databases,'' \emph{ACM TODS}, vol.~34, no.~2, p.~11, 2009.

\bibitem{chairunnanda2014confluxdb}
P.~Chairunnanda, K.~Daudjee, and M.~T. {\"O}zsu, ``Confluxdb: multi-master
  replication for partitioned snapshot isolation databases,'' \emph{PVLDB},
  vol.~7, no.~11, pp. 947--958, 2014.

\bibitem{bailis2014scalable}
P.~Bailis, A.~Fekete, J.~M. Hellerstein, A.~Ghodsi, and I.~Stoica, ``Scalable
  atomic visibility with ramp transactions,'' in \emph{SIGMOD}, 2014, pp.
  27--38.

\bibitem{bailis2015coordination}
P.~D. Bailis, ``Coordination avoidance in distributed databases,'' Ph.D.
  dissertation, University of California, Berkeley, 2015.

\bibitem{adya1999weak}
A.~Adya, ``Weak consistency: a generalized theory and optimistic
  implementations for distributed transactions,'' Ph.D. dissertation,
  Massachusetts Institute of Technology, 1999.

\bibitem{fekete2005making}
A.~Fekete, D.~Liarokapis, E.~O'Neil, P.~O'Neil, and D.~Shasha, ``Making
  snapshot isolation serializable,'' \emph{ACM TODS}, vol.~30, no.~2, pp.
  492--528, 2005.

\bibitem{tu2013speedy}
S.~Tu, W.~Zheng, E.~Kohler, B.~Liskov, and S.~Madden, ``Speedy transactions in
  multicore in-memory databases,'' in \emph{SOSP}, 2013, pp. 18--32.

\bibitem{Cahill2008}
M.~J. Cahill, U.~R\"{o}hm, and A.~D. Fekete, ``Serializable isolation for
  snapshot databases,'' in \emph{SIGMOD}, 2008, pp. 729--738.

\end{thebibliography}

\end{document}